\documentclass[11pt,a4paper]{article}
\usepackage{jheppub}
\usepackage{amsmath,amsfonts,amsthm,graphicx,ulem}

\newcommand{\comment}[1]{}
\newcommand{\bal}{\begin{align}}
\newcommand{\eal}{\end{align}}
\newcommand{\beq}{\begin{equation}}
\newcommand{\eeq}{\end{equation}}
\newcommand\beqa{\begin{eqnarray}}
\newcommand\eeqa{\end{eqnarray}}
\newcommand\bea{\begin{array}}
\newcommand\eea{\end{array}}

\renewcommand{\geq}{\geqslant}

    \newcommand{\nn}{\nonumber}
    
    \newcommand{\COMMENT}[1]{}
    
    \newcommand{\neqa}{\nonumber\end{eqnarray}}

\def\a{{\alpha}}

\def\[{\left[}
\def\]{\right]}

\def\l{\lambda}

\def\a{\alpha}
\def\b{\beta}

\def\<{\langle}
\def\>{\rangle}

\def\Tr{\text{Tr}~}

\def\i2{\frac{i}{2}}

\def\p{\partial}
\def\cO{{\cal O}}

\title{ \center{Three-point correlators of twist-2 operators}\\ \center{ in N=4 SYM at Born approximation}
}
\author[a,b,1]{Vladimir Kazakov%
\note{member of Institut Universitaire de France}}
\author[a]{Evgeny Sobko}

\affiliation[a]{Ecole Normale Superieure, LPT,  75231 Paris CEDEX-5,
  France}
\affiliation[b]{Universit\'e Paris-VI, Paris, France}

\emailAdd{evgenysobko AT gmail.com}
\emailAdd{kazakov AT lpt.ens.fr}

\abstract{We calculate  two different types of  3-point correlators involving  twist-2 operators in the leading weak coupling approximation and all orders in \(N_c\) in N=4 SYM theory. Each of three operators in the first correlator can be any component of twist-2 supermultiplet, though the explicit calculation was done for a particular component which is an \(SU(4)\) singlet. It is calculated in the leading, Born approximation for arbitrary spins  \(j_1,j_2,j_3\). The result  significantly simplifies  when at least one of the spins is large or  equal to zero and the coordinates are restricted to the 2d plane spanned by two light-rays. The second correlator involves two twist-2 operators  \(\text{Tr} (X\nabla^{j_1}X)+...\), \(\text{Tr} (Z\nabla^{j_2}Z)+...\)  and one Konishi operator  \(\text{Tr} [\bar Z,\bar X]^2\). It vanishes in the lowest \(g^{0}\) order and is computed at the leading \(g^2\) approximation.
}
\subheader{LPT ENS-12/47}
\usepackage{varioref}
\usepackage{makeidx}
\makeindex
\usepackage{amssymb}

\begin{document}

  \maketitle

  \section{Introduction}
The operator product expansion in  N=4 SYM theory, as in any CFT, is completely characterized by its 2-point and 3-point correlators, or, in other words, by the spectrum of anomalous dimensions \(\Delta_j(\l)\) of local conformal operators \(\cO_j(x)\)
and by the structure constants \(C_{ijk}(\l)\). \footnote{which are tensors in the general case of operators with spin}.
Both the dimensions and the structure constants are in general complicated functions of  SYM coupling \(\l\) and of  quantum numbers of the operators. Their calculation is a difficult task generally achievable only within a few orders of Feynman perturbation theory w.r.t. \(\l=g^2N\). Things look much better in the planar 't~Hooft limit. In this case, not only we can efficiently treat the theory in the large coupling limit, due to the AdS/CFT correspondence, but  also there is a strong evidence that the theory is integrable at any coupling \(\l\) \cite{Beisert:2006ez,Gromov:2009tv}. This integrability  allows, at least in principle and often in practice,    efficient calculations of anomalous dimensions at arbitrary \(\l\) via the spectral Y-system \cite{Gromov:2009tv,Gromov:2011cx} (or its TBA version \cite{Bombardelli:2009ns,Gromov:2009bc,Arutyunov:2009ur}),  either
numerically  \cite{Gromov:2009zb,Gromov:2011de,Frolov:2012zv,Gromov:2011cx}
or in the  strong and weak coupling expansion (up to 3 loops in strong coupling \cite{Gromov:2011bz}  and up to 8 loops in weak coupling in  \cite{Leurent:2012ab,Bajnok:2012bz,Leurent:2013mr}).

The situation with the structure constants is more complicated, even in the planar limit. For the moment, we have no closed equations, similar to the spectral AdS/CFT Y-system, defining  \(C_{ijk}(\l)\)  for any coupling. One exceptional case is the  correlators of BPS-operators which are known at any coupling \(\lambda\) \cite{Lee:1998bxa,Freedman:1998tz,Arutyunov:1999en,D'Hoker:1999ea}.  The study for more general operators is limited to the case by case computations in the strong coupling limit  \cite{Zarembo:2010rr,Janik:2011bd,Kazama:2012is,Costa:2010rz,Buchbinder:2011jr,Russo:2010bt,Bissi:2011dc,Caputa:2012yj}.   In the  weak  coupling limit, a few interesting particular cases are computed, sometimes up to 3 loops  \cite{Okuyama:2004bd,Roiban:2004va,Alday:2005nd,Plefka:2012rd,Engelund:2012re,Georgiou:2012zj,Eden:2011we,Eden:2012rr}, and in the \(su(2)\)  sector the result is known in general up to one loop  \cite{Gromov:2012vu,Kostov:2012yq}, in a closed form, due to the extensive use of integrability.  However, this study shows a presence of some integrable structures giving a hope for  general solution of this problem, for any \(\l\) and any type of operators. Such
solution could be a generalization  of  the spectral Y-system though it is
probably much more complicated than the latter. At present, we have to continue studying the 3-point functions of various sets of operators in various limits, to acquire more of experience in preparation for attacking the general problem.

In the weak coupling limit, a lot can be done in the leading, Born  approximation or sometimes even in the next few orders. If the operators are short we don't even need to appeal to the integrability for that. For arbitrary operators,  one has to use the integrability to find the  right conformal operators (playing the role of  Bethe wave functions) and to compute the one-loop graph combinatorics.   Most of such results  concern  the closed scalar \(su(2)\)  sector, but  the papers  \cite{Plefka:2012rd,Costa:2012cb}  contains an interesting example involving  twist-2 operators.

In this paper, we calculate two other interesting types of 3-point correlators,
and of the corresponding structure constants, involving  twist-2 operators. The full supermultiplet of twist-2 operators was constructed in  \cite{Belitsky:2003sh}.  Our first  correlator involves  three components of twist-2 super-multiplet with spin \(j\)
and is calculated in the leading approximation for three such operators of arbitrary spins  \(j_1,j_2,j_3\). The result is given by a double integral of three Gegenbauer polynomials, or
a triple sum of elementary functions.  It appears to be very explicit in the case when at least one spin of \(j_1,j_2,j_3\) is large or zero, which can be compared  to the strong coupling computations
of \cite{Kazama:2012is}. The second correlator involves two twist-2 operators  \(\text{Tr} (Z\nabla^{j}Z)\)  and one Konishi operator  \(\text{Tr} [\bar Z,\bar X]^2\). It appears to be zero in the  \(g^{0}\) order  and is computed here at the first non-vanishing \(g^2\) order which is in this case the  Born approximation. A computational advantage for this correlator is  the absence of mixing of all three operators with other operators at this order.

In conclusions, we discuss a few lessons to be retained from these calculation, for the efforts to guess more general structures of the operator product expansion in N=4 SYM, as well as some future directions.

  \section{ 3-point correlator of twist-2 operators in the leading approximation }

We start with the computation at the leading order of 3-point correlators of  particular twist-2 operators, belonging to the \( \mathcal{S}^1_{jl}\) component of twist-2 super-multiplet  \cite{Belitsky:2003sh}:

\begin{equation}
\mathcal{S}^1_{jl}=6\mathcal{O}^{gg}_{jl}+\frac{j}{4}\mathcal{O}^{qq}_{jl}+\frac{j(j+1)}{4}\mathcal{O}^{ss}_{jl}\,.
\end{equation}
The operators

\begin{eqnarray}\label{mult-1}
\mathcal{O}^{gg}_{jl}&=&\frac{1}{2}\sigma_j \Tr i^{l-j}( D_{x_2} +D_{x_1})^{l-j}\mathcal{G}^{\frac{5}{2}}_{j-1,x_1,x_2} F^{+\mu}_{\ \ \bot}(x_1)g^{\bot}_{\mu\nu}F^{\nu+}_{\bot}(x_2)|_{x_1=x_2},\label{O^gg}\\
\mathcal{O}^{qq}_{jl}&=& \sigma_j \Tr i^{l-j}( D_{x_2} +D_{x_1})^{l-j}\mathcal{G}^{\frac{3}{2}}_{j,x_1,x_2} \bar{\lambda}_{\dot{\alpha}A}\sigma^{+\dot{\alpha}\beta}(x_1)\lambda^A_\beta(x_2)|_{x_1=x_2},\label{O^qq} \label{mult-2}\\
\mathcal{O}^{ss}_{jl}&=&  \frac{1}{2}\sigma_j \Tr i^{l-j}( D_{x_2} +D_{x_1})^{l-j}\mathcal{G}^{\frac{1}{2}}_{j+1,x_1,x_2} \bar{\phi}_{AB}(x_1)\phi^{AB}(x_2)|_{x_1=x_2},\label{O^ss}
\label{mult-3}\end{eqnarray}
realize the highest-weight representation of \(sl(2,R)\) at the \(g^0 \) order   when  \(j=l\), and represent its descendants in case of other possible values of \(l\).  We have introduced
the  differential operator \(\mathcal{G}^{\alpha}_{n,x_1,x_2}=i^n( D_{x_2} +D_{x_1})^n C_n^{\alpha}(\frac{D_{x_2}-D_{x_1}}{D_{x_2}+D_{x_1}})\), where \(C_n^{\alpha}(x)\) - Gegenbauer polynomial of order \(n\) with index \(\alpha\). \(D_x\) are covariant derivatives in the light-like direction \(n_+\): \(D_x=n_+^\mu(\partial_\mu-igA_\mu)=\partial_+-igA_+\) and \(\sigma_j=1-(-1)^j\) (operators at \(g^0 \) order   are defined only for odd \(j\)). Also we will use the operator \(\mathcal{G}^{\alpha; g=0}_{n,x_1,x_2}\) , which is given by the same expression as \(\mathcal{G}^{\alpha}_{n,x_1,x_2}\), but with the coupling constant \(g=0\) which is equivalent to the replacement of covariant derivatives by the ordinary ones. For other notations and definition of fields see appendix \ref{AppNotations}.  Acting on a correlation function of three primary operators by derivatives, we can get correlators of any descendants. Thus, we can restrict our attention to the case of three primary operators with \(j=l\) without the loss of generality. These superconformal primary operators \(S^1_{jj}\) have the one-loop anomalous dimension \(\Delta= 3+j+\frac{g^2N_c}{2\pi^2}(\psi(j+2)-\psi(1)) \). In \(g^0 \) order   calculation we should contract only the fields of the same type, and the calculation of correlator \(\langle\mathcal{S}^1_{j_1j_1}(x)\mathcal{S}^1_{j_2j_2}(y)\mathcal{S}^1_{j_3j_3}(z)\rangle\) reduces to the calculation of the sum of three independent correlators:

\begin{gather}
\langle\mathcal{S}^1_{j_1}(x)\mathcal{S}^1_{j_2}(y)\mathcal{S}^1_{j_3}(z)\rangle=6^3\langle \mathcal{O}^{gg}_{j_1}(x)\mathcal{O}^{gg}_{j_2}(y)\mathcal{O}^{gg}_{j_3}(z) \rangle+
\frac{j_1 j_2 j_3}{4^3}\langle \mathcal{O}^{qq}_{j_1}(x) \mathcal{O}^{qq}_{j_2}(y) \mathcal{O}^{qq}_{j_3}(z) \rangle + \notag \\
+\frac{j_1(j_1+1)j_2(j_2+1)j_3(j_3+1)}{4^3}\langle \mathcal{O}^{ss}_{j_1}(x)\mathcal{O}^{ss}_{j_2}(y)\mathcal{O}^{ss}_{j_3}(z)\rangle, \label{3pTw2TrL}
\end{gather}
where we have introduced short notations \(\mathcal{S}^1_{j}=\mathcal{S}^1_{jj}\) and \(\mathcal{O}^{xx}_{j}=\mathcal{O}^{xx}_{jj}\).

\subsection{2 point function}

Let us start with the \(g^0 \) order   calculation of the 2-point function of these twist-2 operators which will be needed anyway for the normalization of their 3-point correlator, in order to extract the corresponding structure constants. The two point correlation function of conformal primary operators \(O^{j_1,...,j_k}\) has a fixed tensor structure:
\begin{gather}
\langle O^{\mu_1,...,\mu_k}(x)\bar{O}^{\nu_1,...,\nu_k}(y)\rangle=C_{O\bar{O}}\frac{I^{\mu_1\nu_1}...I^{\mu_k\nu_k}}{|x-y|^{2\Delta}}, \\
I^{\mu\nu}= g^{\mu\nu}-\frac{2(x-y)^\mu(x-y)^\nu}{|x-y|^2}.
\end{gather}
Due to the fact that all indexes are contracted with the light-like vector \(n_+\) in our case, we get the following formula for the operator with \(k\) \(+\)indices:

\begin{equation}
\langle O_{+,...,+}(x)\bar{O}_{+,...,+}(y)\rangle=C_{O\bar{O}} \frac{(-2(x-y)^2_+)^k}{|x-y|^{2(\Delta+k)}}.
\end{equation}

It is clear that  the restriction of coordinates \(x\) and \(y\) to 2d plane \(\{\bot\}\) does not lead to a loss of generality in case of the two-point correlation functions.

\par\medskip

At the leading order, we can factor out all the derivatives from the quantum average, and first calculate the gaussian integrals over fields. It gives

\begin{gather}
\langle\mathcal{O}^{gg}_{j_1}(x) \mathcal{O}^{gg}_{j_2}(y)\rangle=\notag\\
=\mathcal{N}^2\frac{\sigma_{j_1}\sigma_{j_2}}{4}\mathcal{G}^{\frac{5}{2};g=0}_{j_1-1,x_1,x_2}\enskip\mathcal{G}^{\frac{5}{2};g=0}_{j_2-1,y_1,y_2}\langle \Tr\left( F^{+\mu}_{\ \ \bot}(x_1)g^{\bot}_{\mu\nu}F^{\nu+}_{\bot}(x_2)\right)\Tr\left(
F^{+\mu'}_{\ \ \bot}(y_1)g^{\bot}_{\mu'\nu'}F^{\nu'+}_{\bot}(y_2)\right) \rangle|_{\substack{x_1=x_2\\
y_1=y_2}}=\notag\\
=\mathcal{N}^2\sigma_{j_1}\sigma_{j_2}(N_c^2-1)\,\mathcal{G}^{5/2;g=0}_{j_1-1,x_1,x_2}\enskip\mathcal{G}^{5/2;g=0}_{j_2-1,y_1,y_2}\partial_{x_1+}\partial_{x_2+}\partial_{y_1+}\partial_{y_2+}\frac{1}{|x_1-y_1|^2|x_2-y_2|^2}|_{\substack{x_1=x_2\\
y_1=y_2}},
\end{gather}
where in the last equality we used the evenness of Gegenbauer polynomials. The factor \(\mathcal{N}=-\frac{1}{8\pi^2}\) is just a normalization for two-point propagators (see appendix \ref{AppNotations}).
As it was noticed before, we can put \((x-y)_{\bot}=0\). By ordering the points as \(x_->y_-\) and using the  formula\(\frac{1}{(x-y)^k_-}=\int\limits_{0}^{\infty}ds \frac{s^{k-1}}{\Gamma(k)}e^{-s (x-y)_-}\), we can rewrite the action of differential operators as an integral and carry out the calculation (see appendix \ref{AppGeg}). Finally we obtain:
\begin{gather}
\langle\mathcal{O}^{gg}_{j_1}(x) \mathcal{O}^{gg}_{j_2}(y)\rangle_\natural=\delta_{j_1,j_2}\mathcal{N}^2\sigma_{j_1}^2(N_c^2-1)\frac{\Gamma(2j_1+4)\Gamma(j_1+4)}{2^73^2\Gamma(j_1)(j_1+3/2)}\frac{1}{(x-y)_+^2(x-y)_-^{2j_1+4}},
\end{gather}
where the symbol "\(\natural\)" here and below means the restriction to the plane spanned by \(\{n_+,n_-\}\). The full 4-d answer reads as follows:
\begin{gather}
\langle\mathcal{O}^{gg}_{j_1}(x) \mathcal{O}^{gg}_{j_2}(y)\rangle=\delta_{j_1,j_2}\mathcal{N}^2\sigma_{j_1}^2(N_c^2-1)\frac{\Gamma(2j_1+4)\Gamma(j_1+4)2^{2j_1-3}}{3^2\Gamma(j_1)(j_1+3/2)}
\frac{(x-y)_+^{2j_1+2}}{((x-y)^2)^{2j_1+4}}.
\end{gather}
Similarly we get the following leading order correlators for the operators \(\mathcal{O}^{qq}_{j}\) and \(\mathcal{O}^{ss}_{j}\):
\begin{gather}
\langle \mathcal{O}^{qq}_{j_1}(x)\mathcal{O}^{qq}_{j_2}(y)\rangle=\delta_{j_1,j_2}\mathcal{N}^2\sigma_{j_1}^2(N_c^2-1)\frac{\Gamma(2j_1+4)\Gamma(j_1+3)2^{2j_1+3}}{\Gamma(j_1+1)(j_1+3/2)}
\frac{(x-y)_+^{2j_1+2}}{((x-y)^2)^{2j_1+4}},\\
\langle \mathcal{O}^{ss}_{j_1}(x)\mathcal{O}^{ss}_{j_2}(y)\rangle=\delta_{j_1,j_2}\mathcal{N}^2\sigma_{j_1}^2(N_c^2-1)\frac{3\Gamma(2j_1+4)2^{2j_1-1}}{j_1+3/2}\frac{(x-y)_+^{2j_1+2}}{((x-y)^2)^{2j_1+4}},
\end{gather}

And finally one can assemble the expression for the \(\langle\mathcal{S}^1_{j_1}(x)\mathcal{S}^1_{j_2}(y)\rangle\) :
\begin{gather}
\langle\mathcal{S}^1_{j_1}(x)\mathcal{S}^1_{j_2}(y)\rangle=\delta_{j_1j_2}\sigma_{j_1}^2(N_c^2-1)H(j_1)\frac{(x-y)_+^{2j_1+2}}{((x-y)^2)^{2j_1+4}},\\
H(j_1)=\mathcal{N}^2j_1(j_1+1)(96+115j_1+35j_1^2)2^{2j_1-4}(2j_1+2)!.
\end{gather}
We will also use the two-point correlator \(\langle\mathcal{S}^1_{j_1}(x)\mathcal{S}^1_{j_2}(y)\rangle_{\natural}\) with coordinates \(x,y\)  restricted to the \(\{n_+,n_-\}\) 2d-plane which is equal to

\begin{gather}
\langle\mathcal{S}^1_{j_1}(x)\mathcal{S}^1_{j_2}(y)\rangle_{\natural}=\delta_{j_1j_2}H_{\natural}(j_1)\frac{1}{(x-y)^2_+(x-y)_-^{2j_1+4}},\ \ H_{\natural}(j_1)=\sigma_{j_1}^2(N_c^2-1)\frac{H(j_1)}{2^{2j_1+4}}.\label{NormalizationOfTwoS}
\end{gather}

Asymptotic form of \(H_{\natural}(j_1)\) when \(j_1\rightarrow \infty\) reads as follows:

\begin{gather}\label{ApprH}
H_{\natural}(j_1)=\mathcal{N}^2\sigma_{j_1}^2(N_c^2-1)\frac{35}{64}j_1^6\Gamma(2j_1+1)(1+o(j_1^{-1})).
\end{gather}

\subsection{3-point function}
Let us now proceed  with the calculation of the 3-point correlators.

As was noticed before, the correlation function of three components \(\mathcal{S}^1_{j_1}(x)\) of twist-2 supermultiplet can be reduced to the sum of correlators of operators with the same fields. Direct use of  the Wick rule for the explicit expressions
\eqref{mult-1},\eqref{mult-2},\eqref{mult-3} gives

\begin{eqnarray}
\langle \mathcal{O}^{gg}_{j_1}(x)\mathcal{O}^{gg}_{j_2}(y)\mathcal{O}^{gg}_{j_3}(z) \rangle&=&-2^{10}\mathcal{N}^3(N_c^2-1)\ \mathcal{G}^{\frac{5}{2};g=0}_{j_1-1,x_1,x_2}\enskip\mathcal{G}^{\frac{5}{2};g=0}_{j_2-1,y_1,y_2}\enskip\mathcal{G}^{\frac{5}{2};g=0}_{j_3-1,z_1,z_2}
\left.\Upsilon_2\right|_\clubsuit,\\
\langle \mathcal{O}^{qq}_{j_1}(x) \mathcal{O}^{qq}_{j_2}(y) \mathcal{O}^{qq}_{j_3}(z) \rangle&=&-i2^9\mathcal{N}^3(N_c^2-1)\ \mathcal{G}^{\frac{3}{2};g=0}_{j_1,x_1,x_2}\enskip\mathcal{G}^{\frac{3}{2};g=0}_{j_2,y_1,y_2}\enskip\mathcal{G}^{\frac{3}{2};g=0}_{j_3,z_1,z_2}
\Upsilon_1|_\clubsuit, \\
\langle \mathcal{O}^{ss}_{j_1}(x)\mathcal{O}^{ss}_{j_2}(y)\mathcal{O}^{ss}_{j_3}(z)&=&2^4 3\mathcal{N}^3(N_c^2-1)\ \mathcal{G}^{\frac{1}{2};g=0}_{j_1+1,x_1,x_2}\enskip\mathcal{G}^{\frac{1}{2};g=0}_{j_2+1,y_1,y_2}\enskip\mathcal{G}^{\frac{1}{2};g=0}_{j_3+1,z_1,z_2}
\Upsilon_0|_\clubsuit,
\end{eqnarray}
where we have introduced function
\begin{gather}
\Upsilon_k=\sigma_{j_1}\sigma_{j_2}\sigma_{j_3}\frac{(x_1-y_2)^k_+(y_1-z_2)^k_+(z_1-x_2)^k_+}{(|x_1-y_2|^2)^{1+k}(|y_1-z_2|^2)^{1+k}(|z_1-x_2|^2)^{1+k}}
\end{gather} and a new notation \(\clubsuit=\{x_i=x,\ y_i=y,\ z_i=z\}\), for the sake of brevity.
Now using the formulas (\ref{G12prod})-(\ref{G52prod}) one can get explicit expressions for correlators:

\begin{eqnarray}
\langle \mathcal{O}^{gg}_{j_1}(x)\mathcal{O}^{gg}_{j_2}(y)\mathcal{O}^{gg}_{j_3}(z)\rangle &=&\mathcal{N}_{j_1j_2j_3}2^{-2}3^{-3}j_1(j_1+1)j_2(j_2+1)j_3(j_3+1)\times\notag\\
&\times&\sum\limits_{k_1=0}^{j_1-1}\sum\limits_{k_2=0}^{j_2-1}\sum\limits_{k_3=0}^{j_3-1}\eta_g(k_1,k_2,k_3)\theta(k_1,k_2,k_3),\label{Exactggg}
\end{eqnarray}
 where
\begin{eqnarray*}
\eta_g(k_1,k_2,k_3)&=&\binom{j_1-1}{k_1}\binom{j_1+3}{k_1+2}\binom{j_2-1}{k_2}\binom{j_2+3}{k_2+2}\binom{j_3-1}{k_3}\binom{j_3+3}{k_3+2}\times\\ &\times&(j_1+1-k_1+k_2)!(j_2+1-k_2+k_3)!(j_3+1-k_3+k_1)!
\end{eqnarray*}
and \(\mathcal{N}_{j_1j_2j_3}=\mathcal{N}^3(N_c^2-1)\sigma_{j_1}\sigma_{j_2}\sigma_{j_3}i^{j_1+j_2+j_3+3}2^{j_1+j_2+j_3}\)
\begin{eqnarray}
\langle \mathcal{O}^{qq}_{j_1}(x)\mathcal{O}^{qq}_{j_2}(y)\mathcal{O}^{qq}_{j_3}(z)\rangle&=&\mathcal{N}_{j_1j_2j_3}2^6(j_1+1)(j_2+1)(j_3+1)\times\notag\\
&\times&\sum\limits_{k_1=0}^{j_1}\sum\limits_{k_2=0}^{j_2}\sum\limits_{k_3=0}^{j_3}\eta_q(k_1,k_2,k_3)\theta(k_1,k_2,k_3),
\end{eqnarray}
 where \begin{eqnarray*}\eta_q(k_1,k_2,k_3)&=&\binom{j_1}{k_1}\binom{j_1+2}{k_1+1}\binom{j_2}{k_2}\binom{j_2+2}{k_2+1}\binom{j_3}{k_3}\binom{j_3+2}{k_3+1}\times\\ &\times&(j_1+1-k_1+k_2)!(j_2+1-k_2+k_3)!(j_3+1-k_3+k_1)!\,,\end{eqnarray*}

\begin{gather}
\langle \mathcal{O}^{ss}_{j_1}(x)\mathcal{O}^{ss}_{j_2}(y)\mathcal{O}^{ss}_{j_3}(z)\rangle=\mathcal{N}_{j_1j_2j_3}2^73\sum\limits_{k_1=0}^{j_1+1}\sum\limits_{k_2=0}^{j_2+1}\sum\limits_{k_3=0}^{j_3+1}\eta_s(k_1,k_2,k_3)\theta(k_1,k_2,k_3),\label{Exactsss}
\end{gather}
where
\begin{eqnarray*}
\eta_s(k_1,k_2,k_3)&=&\binom{j_1+1}{k_1}^2\binom{j_2+1}{k_2}^2\binom{j_3+1}{k_3}^2\times\\&\times&(j_1+1-k_1+k_2)!(j_2+1-k_2+k_3)!(j_3+1-k_3+k_1)!
\end{eqnarray*}  and \begin{equation}\theta(k_1,k_2,k_3)=\frac{(x-y)^{j_1+1-k_1+k_2}_+}{(|x-y|^2)^{j_1+2-k_1+k_2}}
\frac{(y-z)^{j_2+1-k_2+k_3}_+}{(|y-z|^2)^{j_2+2-k_2+k_3}}\frac{(z-x)^{j_3+1-k_3+k_1}_+}{(|z-x|^2)^{j_3+2-k_3+k_1}}.
\end{equation}
Now one can assemble the correlator \(\langle\mathcal{S}^1_{j_1}(x)\mathcal{S}^1_{j_2}(y)\mathcal{S}^1_{j_3}(z)\rangle\) using  (\ref{3pTw2TrL}).

General three-point correlation function of operators with spins in CFT contains a sum over the different tensor structures, as in (\ref{SumTens}). Using the formulas (\ref{Exactggg})-(\ref{Exactsss}) one can obtain the expression for an arbitrary tensor structure. But now let us consider a special case by restricting all 3 coordinates to  a special 2d plane \(\{n_+,n_-\}\). This collapses all tensor structures to a single one (see \ref{3pRC}).  In this case the correlator reads as follows
\begin{gather}
\langle O^{\alpha\alpha}_{j_1}(x)O^{\alpha\alpha}_{j_2}(y)O^{\alpha\alpha}_{j_3}(z)\rangle_{\natural}=\notag\\
\frac{B^\alpha_{\natural j_1j_2j_3}}{(x-y)_+(y-z)_+(z-x)_+(x-y)_-^{j_1+j_2-j_3+2}(y-z)_-^{j_2+j_3-j_1+2}(z-x)_-^{j_1+j_3-j_2+2}},
\end{gather}
where \(\alpha\in\{g,q,s\}\)  labels the 3 components of the multiplet of twist-2 operators. Setting, without
a loss of generality, \(x_-=1, y_-=0, z_-=-1\) one can get  \(B^\alpha_{\natural j_1j_2j_3}\) from the above formulas as  finite triple sums:

\begin{eqnarray}
B^g_{\natural j_1j_2j_3}&=&\mathcal{N}_{j_1j_2j_3}i^22^{-2j_2-6}3^{-3}j_1(j_1+1)j_2(j_2+1)j_3(j_3+1)\times\notag\\
&\times&\sum\limits_{k_1=0}^{j_1-1}\sum\limits_{k_2=0}^{j_2-1}\sum\limits_{k_3=0}^{j_3-1}\frac{\eta_g(k_1,k_2,k_3)}{(-2)^{j_3+2-k_3+k_1}},\label{3G2d}
\end{eqnarray}

\begin{gather}
B^q_{\natural j_1j_2j_3}=\mathcal{N}_{j_1j_2j_3}i^22^{-2j_2+2}(j_1+1)(j_2+1)(j_3+1)\sum\limits_{k_1=0}^{j_1}\sum\limits_{k_2=0}^{j_2}\sum\limits_{k_3=0}^{j_3}\frac{\eta_q(k_1,k_2,k_3)}{(-2)^{j_3+2-k_3+k_1}},\label{3Q2d}
\end{gather}

\begin{gather}
B^s_{\natural j_1j_2j_3}=3\mathcal{N}_{j_1j_2j_3}i^22^{-2j_2+3}\sum\limits_{k_1=0}^{j_1+1}\sum\limits_{k_2=0}^{j_2+1}\sum\limits_{k_3=0}^{j_3+1}\frac{\eta_s(k_1,k_2,k_3)}{(-2)^{j_3+2-k_3+k_1}}.\label{3S2d}
\end{gather}
Using (\ref{NormalizationOfTwoS}) we can normalize the  3-point correlator \( \langle\mathcal{S}^1_{j_1}(x_1)\mathcal{S}^1_{j_2}(x_2)\mathcal{S}^1_{j_3}(x_3)\rangle_{\natural}\)  using the 2-point correlators as follows

\begin{eqnarray}
C_{\natural j_1j_2j_3}=\langle\mathcal{S}^1_{j_1}(x_1)\mathcal{S}^1_{j_2}(x_2)\mathcal{S}^1_{j_3}(x_3)\rangle_{\natural}
\sqrt{\prod\limits_{k=1}^3\frac{\langle\mathcal{S}^1_{j_k}(x_{k+1})\mathcal{S}^1_{j_k}(x_{k+2})\rangle_{\natural}}
{\langle\mathcal{S}^1_{j_k}(x_{k})\mathcal{S}^1_{j_k}(x_{k+1})\rangle_\natural\langle\mathcal{S}^1_{j_k}(x_{k})\mathcal{S}^1_{j_k}(x_{k+2})\rangle_{\natural}}}
\end{eqnarray}
which finally gives the structure constant of these twist two operators:

\begin{gather}
C_{\natural j_1j_2j_3}=\frac{\sigma_{j_1}\sigma_{j_2}\sigma_{j_3}}{8}\frac{\left(6^3B^g_{\natural j_1j_2j_3}+\frac{j_1j_2j_3}{4^3}B^q_{\natural j_1j_2j_3}+\frac{j_1(j_1+1)j_2(j_2+1)j_3(j_3+1)}{4^3}B^s_{\natural j_1j_2j_3}\right)}{\sqrt{H_{\natural}(j_1)H_{\natural}(j_2)H_{\natural}(j_3)}}.\label{3pNormalization}
\end{gather}

Expressions for \(B_{\natural j_1j_2j_3}\) dramatically simplify in the case when one of spins equals two. For \(j_3=1\) we get:

\begin{eqnarray}
B^g_{\natural j_1j_21}&=&2\mathcal{N}^3(N_c^2-1)\sigma_{j_1}\sigma_{j_2}i^{j_1+j_2}C_{j_1-1}^{\frac{5}{2}}(1)C_{j_2-1}^{\frac{5}{2}}(1)(j_1+j_2)!,\\
B^q_{\natural j_1j_21}&=&-2^43\mathcal{N}^3(N_c^2-1)\sigma_{j_1}\sigma_{j_2}i^{j_1+j_2}C^{\frac{3}{2}}_{j_1}(1)C^{\frac{3}{2}}_{j_2}(1)\times\notag\\
&\times&(2+3(j_1+j_2)+j_1^2+j_2^2)(j_1+j_2)!,\\
B^s_{\natural j_1j_21}&=&2^23\mathcal{N}^3\sigma_{j_1}\sigma_{j_2}i^{j_1+j_2}(N_c^2-1)(j_1+j_2)![16+30(j_1+j_2)+19(j_1^2+j_2^2)+\notag\\
&+&6(j_1^3+j_2^3)+j_1^4+j_2^4+36j_1j_2+4j_1^2j_2^2+12(j_1^2j_2+j_1j_2^2)],
\end{eqnarray}
where  \(C^\alpha_{n}(1)=\frac{\Gamma(2\alpha+n)}{\Gamma(2\alpha)\Gamma(n+1)}\) is the value of Gegenbauer polynomial at the argument equal 1.
\subsection{Asymptotic form of 3-point functions}
\subsubsection{The case of three large spins.}
Another case when expressions (\ref{3G2d})-(\ref{3S2d}) drastically simplify is the limit of large spins \(j_1\sim j_2\sim j_3\gg 1\). The case when both the spin and the twist of two operators are large and the third operator is BPS was investigated in \cite{Georgiou:2011qk} and the agreement between weak and strong coupling was found. In our case we can apply the saddle-point approximation.   First, using the Euler-Maclaurin formula  we  approximate the triple sum through the integral. Using Stirling approximation for binomial coefficients in the function \(\eta\) we get for  (\ref{3G2d})-(\ref{3S2d}) the following integral representation:

\begin{gather}
B^{\alpha}_{\natural j_1,j_2,j_3}\simeq \int\int\int dk_1dk_2dk_3 f_{\alpha}(j_1,j_2,j_3,k_1,k_2,k_3)e^S,
\end{gather}
where the label \(\alpha\) denotes the type of correlator, \(f_{\alpha}(j_1,j_2,j_3,k_1,k_2,k_3)\) is a rational function of a finite order in variables \(k_i\) and the action \(S\) in the exponent is the universal function for all correlators given by

\begin{gather}
S=S(j_1,j_2,j_3,k_1,k_2,k_3)=(-k_1+k_3)\ln(-2)-2k_1\ln k_1-\notag\\
-2(j_1-k_1)\ln (j_1-k_1)-2k_2\ln k_2-2(j_2-k_2)\ln (j_2-k_2)-2k_3\ln k_3-\notag\\
-2(j_3-k_3)\ln (j_3-k_3)+(j_1-k_1+k_2)\ln (j_1-k_1+k_2)+\notag\\
+(j_2-k_2+k_3)\ln (j_2-k_2+k_3)+(j_3-k_3+k_1)\ln (j_3-k_3+k_1).\label{SPaction}
\end{gather}
The 3 saddle-point equations \(\frac{\p S}{\p k_j}\) =0,\quad k=1,2,3, read as follows

\begin{gather}
\left\{
\begin{aligned}
-\frac{(j_1-k_1)^2(j_3-k_3+k_1)}{2k_1^2(j_1-k_1+k_2)}&=1,\\
\frac{(j_2-k_2)^2(j_1-k_1+k_2)}{k_2^2(j_2-k_2+k_3)}&=1,\\
-\frac{2(j_3-k_3)^2(j_2-k_2+k_3)}{k_3^2(j_3-k_3+k_1)}&=1.
\end{aligned}
\right.\label{SPeq}
\end{gather}

The solution  rendering the main contribution can be obtained explicitly

\begin{gather}
\left\{
\begin{aligned}
k_1&=\frac{j_1(j_3-j_1)}{j_1+j_2+j_3},\\
k_2&=\frac{j_2(2j_1+j_2)}{2(j_1+j_2+j_3)},\\
k_3&=\frac{j_3(j_2+2j_3)}{j_1+j_2+j_3}.
\end{aligned}
\right.\label{SPsol}
\end{gather}

The function \(S,\) and hence the saddle-point equations (\ref{SPeq}) as well as  their solution, (\ref{SPsol}) have an asymmetry which corresponds to our particular choice of coordinates of 3 points  \(x_-=1, y_-=0, z_-=-1\) in the correlator. However, we finally  get a fully symmetric answer, as it should be:

\begin{gather}
B^g_{\natural j_1j_2j_3}=\frac{1}{3^32^9\sqrt{\pi}}\Lambda(j_1,j_2,j_3)(1+\cO(j_k^{-1})),\\
B^q_{\natural j_1j_2j_3}=-\frac{1}{2\sqrt{\pi}(j_1j_2j_3)}\Lambda(j_1,j_2,j_3)(1+\cO(j_k^{-1})),\\
B^s_{\natural j_1j_2j_3}=\frac{3}{\sqrt{\pi}(j_1j_2j_3)^2}\Lambda(j_1,j_2,j_3)(1+\cO(j_k^{-1})),
\end{gather}
where \(\Lambda(j_1,j_2,j_3)=i^{j_1+j_2+j_3+3}\mathcal{N}^3(N_c^2-1)\sigma_{j_1}\sigma_{j_2}\sigma_{j_3}(j_1j_2j_3)^{\frac{3}{2}}(j_1+j_2+j_3+4)!\).

Using formula (\ref{3pTw2TrL})we get:

\begin{gather}
\langle\mathcal{S}^1_{j_1}(x)\mathcal{S}^1_{j_2}(y)\mathcal{S}^1_{j_3}(z)\rangle_{\natural}=\frac{7}{2^7\sqrt{\pi}}\Lambda(j_1,j_2,j_3)(1+\cO(j_k^{-1})).
\end{gather}
And finally, after normalization (\ref{3pNormalization}) we get the following explicit structure constant of 3  twist-2 operators  (for the chosen component of the multiplet) in the limit \(j_1\sim j_2\sim j_3\to\infty\):

\begin{gather}
C_{\natural j_1j_2j_3}\simeq\sigma_{j_1}\sigma_{j_2}\sigma_{j_3}\frac{1}{\sqrt{N_c^2-1}}\frac{i^{j_1+j_2+j_3+3}}{\pi^{\frac{1}{2}}5^{\frac{3}{2}}7^{\frac{1}{2}}2(j_1j_2j_3)^{\frac{3}{2}}}
\frac{(j_1+j_2+j_3+4)!}{\sqrt{(2j_1)!(2j_2)!(2j_3)!}}(1+\cO(j_k^{-1})).
\end{gather}

\subsubsection{The case of two large and one arbitrary spin }
The similar analysis can be done in the case when two spins are large \(j_1\sim j_2 \gg j_3\). Expressions for \(B_{j_1j_2j_3}\) in this case read as follows:

\begin{eqnarray}
B^g_{\natural j_1j_2j_3}&=&\mathcal{N}^3(N_c^2-1)i^{j_1+j_2+j_3+3}2^{-6}3^{-2}\sigma_{j_1}\sigma_{j_2}\sigma_{j_3}j_1^4j_2^4(j_1-j_2)^{j_3-1}\times\notag\\
&\times&C^{\frac{5}{2}}_{j_3-1}
(\frac{j_1^2+j_2^2}{j_1^2-j_2^2})(j_1+j_2)!(1+\cO(j_k^{-1})),\\
B^q_{\natural j_1j_2j_3}&=&\mathcal{N}^3(N_c^2-1)i^{j_1+j_2+j_3+1}2\sigma_{j_1}\sigma_{j_2}\sigma_{j_3}j_1^2j_2^2(j_1-j_2)^{j_3}\times\notag\\
&\times&C^{\frac{3}{2}}_{j_3}
(\frac{j_1^2+j_2^2}{j_1^2-j_2^2})(j_1+j_2+1)!(1+\cO(j_k^{-1})),\\
B^s_{\natural j_1j_2j_3}&=&\mathcal{N}^3(N_c^2-1)i^{j_1+j_2+j_3+3}6\sigma_{j_1}\sigma_{j_2}\sigma_{j_3}(j_1-j_2)^{j_3+1}\times\notag\\
&\times&C^{\frac{1}{2}}_{j_3+1}
(\frac{j_1^2+j_2^2}{j_1^2-j_2^2})(j_1+j_2+2)!(1+\cO(j_k^{-1})).
\end{eqnarray}
Substituting it into \eqref{3pNormalization} and using the approximation \eqref{ApprH} for large \(j_1\) and \(j_2\)
we obtain the final expression for the structure constants in this limit.

\subsubsection{The case of one large and two arbitrary spins}
In the case of one large spin \(j_1\gg j_2,j_3\) we get:
\begin{eqnarray}
B^g_{\natural j_1j_2j_3}&=&\mathcal{N}^3(N_c^2-1)i^{j_1+j_2+j_3+3}2^{-3}3^{-1}\sigma_{j_1}\sigma_{j_2}\sigma_{j_3}\times\notag\\
&\times&C^{\frac{5}{2}}_{j_2-1}(1)C^{\frac{5}{2}}_{j_3-1}(1)(j_1+j_2+j_3+3)!(1+\cO(j_k^{-1})),\\
B^q_{\natural j_1j_2j_3}&=&\mathcal{N}^3(N_c^2-1)i^{j_1+j_2+j_3+1}4\sigma_{j_1}\sigma_{j_2}\sigma_{j_3}\times\notag\\
&\times&C^{\frac{3}{2}}_{j_2}(1)
C^{\frac{3}{2}}_{j_3}(1)(j_1+j_2+j_3+3)!(1+\cO(j_k^{-1})),\\
B^q_{\natural j_1j_2j_3}&=&\mathcal{N}^3(N_c^2-1)i^{j_1+j_2+j_3+3}6\sigma_{j_1}\sigma_{j_2}\sigma_{j_3}(j_1+j_2+j_3+3)!(1+\cO(j_k^{-1})).
\end{eqnarray}
Substituting it into \eqref{3pNormalization} and using the approximation \eqref{ApprH} for large \(j_1\) we obtain the final expression for the structure constants in this limit.

\section{Correlation function of two twist-2 and one Konishi operators}

In this section we are going to calculate the three-point correlation function of  one Konishi operator
\begin{equation}\label{Konishi}\cO_K=\Tr[\bar{X},\bar{Z}]^2=2\Tr(\bar{X}\bar{Z}\bar{X}\bar{Z})-2\Tr(\bar{X}^2\bar{Z}^2) \end{equation}
   and two scalar twist-2 operators of spins \(j_1+1\) and \(j_2+1\) from the \([O_{jj}^{ss,\ 20}]_{AB}^{CD}\) component of twist-2 supermultiplet \cite{Belitsky:2003sh}:

\begin{gather}
[O_{j_1j_1}^{ss,\ 20}]_{41}^{32}(\alpha)=6\sigma_{j_1}\Tr[\mathcal{G}^{\frac{1}{2}}_{j_1+1,\alpha_1,\alpha_2}X(\alpha_1)X(\alpha_2)]|_{\alpha=\alpha_1=\alpha_2},\notag\\
[O_{j_2j_2}^{ss,\ 20}]_{43}^{21}(\alpha)=6\sigma_{j_2}\Tr[\mathcal{G}^{\frac{1}{2}}_{j_2+1,\alpha_1,\alpha_2}Z(\alpha_1)Z(\alpha_2)]|_{\alpha=\alpha_1=\alpha_2}.
\end{gather}
Let us note that these operators belong to the irrep  \textbf{20} w.r.t the \(SU(4) \) and vanish for even \(j\). Minimal spin of such operators is equal to 2, when \(j=1\). Let us note  that a mixing with double-trace operators could potentially make contribution to such correlators which have only two "sides"\footnote{i.e. the propagators  connecting the  operators, say, \(O_1\), \(O_2\),\(O_3\) go only between \(O_1\) and \(O_2\) or \(O_2\) and \(O_3\), but not between \(O_1\) and \(O_3\), as in Fig\ref{ris:Tree-level}} in the leading order. But it turns out that both Konishi and twist-2  operators do not have such a mixing. It is true in the case of twist-2 operators as they are constructed only  from  two scalar fields, and there is no possibility to construct double-trace operators from them. In its turn, \(\cO_K\) is a descendent of the lowest component of Konishi supermultiplet and it does not mix with other operators \cite{Ryzhov:2001bp,Bellucci:2004ru}.

\subsection{$g^0$ calculation}

 Let us first show that  this correlator is zero at the  \(g^0\) order.
Using the same point-splitting procedure as in the previous section we can rewrite the correlator as follows:

\begin{eqnarray}
&&\langle [O_{j_1j_1}^{ss,\ 20}]_{41}^{32}(\alpha)[O_{j_2j_2}^{ss,\ 20}]_{43}^{21}(\beta) \rangle=\nn\\
&&=\mathcal{K}\langle\text{Tr}\left(X(\a_1)X(\a_2)\right)
\Tr\left( Z(\b_1)Z(\b_2)\right)\Tr[\bar{X},\bar{Z}]^2(\gamma)\rangle|_{\begin{smallmatrix}\alpha_{1,2}=\alpha\\ \beta_{1,2}=\beta\end{smallmatrix}},
\label{f.f.}
\end{eqnarray}
where we have introduced the differential operator \(\mathcal{K}=36\sigma_{j_1}\sigma_{j_2}\ \mathcal{G}^{\frac{1}{2};g=0}_{j_1+1,\alpha_1,\alpha_2}\enskip\mathcal{G}^{\frac{1}{2};g=0}_{j_2+1,\beta_1,\beta_2}\).
The quantum average via Wick rule at this approximation  in (\ref{f.f.}) gives
(omitting for brevity the  coordinate dependence in the l.h.s., obvious from \eqref{f.f.}):

\begin{gather}
\langle\Tr X^2\Tr Z^2\Tr(\bar{X}\bar{Z}\bar{X}\bar{Z})\rangle=\frac{4\mathcal{N}^4(-N_c+\frac{1}{N_c})}{|\alpha_1-\gamma|^2|\alpha_2-\gamma|^2|\beta_1-\gamma|^2|\beta_2-\gamma|^2},\label{a.a.tree.l1}\\
\langle\Tr X^2\Tr Z^2\Tr(\bar{X}^2\bar{Z}^2)\rangle=\frac{4\mathcal{N}^4(N_c^3-2N_c+\frac{1}{N_c})}{|\alpha_1-\gamma|^2|\alpha_2-\gamma|^2|\beta_1-\gamma|^2|\beta_2-\gamma|^2}. \label{a.a.tree.l2}
\end{gather}
Now,  we can apply  the differential operators \(\mathcal{G}^{\frac{1}{2};g=0}_{j_1+1,\alpha_1,\alpha_2}\) and \(\mathcal{G}^{\frac{1}{2};g=0}_{j_2+1,\beta_1,\beta_2}\) to (\ref{a.a.tree.l1}) and (\ref{a.a.tree.l2}),
 using the relations:
\begin{gather}
(a+b)^nC_n^{\frac{1}{2}}(\frac{a-b}{a+b})=\sum \limits_{k=0}^n a^kb^{n-k}(-1)^{n-k}\binom{n}{k}^2,
\\
D_{x_-}^k \frac{1}{|x|^2}=D_{x_-}^k \frac{1}{2x_+x_--x^2_{\perp}}
=\frac{(-1)^kk!(2x_+)^k}{(|x|^2)^{k+1}}\,.
\end{gather}
The  action of \(\mathcal{G}^{\frac{1}{2};g=0}_{j_1+1,\alpha_1,\alpha_2}\) on (\ref{a.a.tree.l1}) or (\ref{a.a.tree.l2}) gives a factor which vanishes after putting \(\alpha_1=\alpha_2=\alpha\)~:

\begin{gather}
\lim_{\a_{1,2}\to \a}(D_{\alpha_1}+D_{\alpha_2})^nC_n^{\frac{1}{2}}\left(\frac{D_{\alpha_1}-D_{\alpha_2}}{D_{\alpha_1}+D_{\alpha_2}} \right)\frac{1}{|\alpha_2-\gamma_3|^2|\alpha_1-\gamma_4|^2}=\notag\\
=\lim_{\a_{1,2}\to \a}\sum\limits_{k=0}^n(-1)^{n-k}(C_n^k)^2 \frac{k!(-2(\alpha_{1+}-\gamma_{4+}))^k}{(|\alpha_1-\gamma_4|^2)^{k+1}}\frac{(n-k)!(-2(\alpha_{2+}-\gamma_{3+}))^{n-k}}{(|\alpha_2-\gamma_3|^2)^{n-k+1}}=
\notag\\=
\frac{n!(2(\alpha_+-\gamma_+))^n}{(|\alpha-\gamma|^2)^{n+2}}\sum\limits_{k=0}^{n}C_n^k (-1)^k=0. \label{0inTreeL}
\end{gather}
Hence finally we conclude that this correlator vanishes at the at the leading order:
\begin{gather}
\langle [O_{j_1j_1}^{ss,\ 20}]_{41}^{32}(\alpha)[O_{j_2j_2}^{ss,\ 20}]_{43}^{21}(\beta)\Tr[X,Z]^2(\gamma) \rangle|_{g=0}=0.
\end{gather}
Before proceeding with the \(g^2\) calculation, we write below the two-point correlation functions of two Konishi operators and  of two twist-2 operators in the leading order  which we will use for the normalization of the  3-point correlator in the leading \(g^2\) approximation:

\begin{figure}
 \center{\includegraphics[scale=0.2]{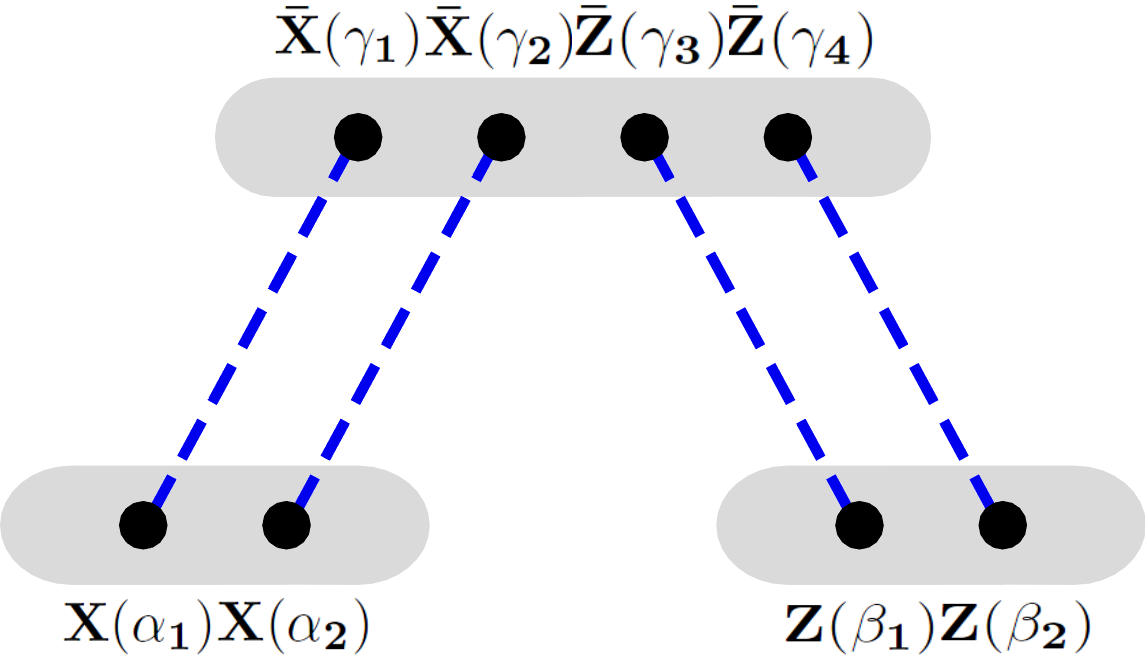}}
 \caption{Leading diagram.}
\label{ris:Tree-level}
\end{figure}

\begin{gather}
\langle\Tr[X,Z]^2(x)\Tr[\bar X,\bar Z]^2(y) \rangle\vert_{ _{g=0}}=12\mathcal{N}^4(N_c^4-N_c^2)\frac{1}{|x-y|^8},\label{KonishiNorm}
\end{gather}

\begin{gather}
\langle [O_{j_1j_1}^{ss,\ 20}]_{41}^{32}(x)\overline{[O_{j_1j_1}^{ss,\ 20}]_{41}^{32}}(y)\rangle\vert_{_{ g=0}}=\delta_{j_1,j_2}\mathcal{N}^2(N_c^2-1)\sigma_{j_1}^2\Gamma(2j_1+3)3^22^{2j_1+5}\frac{(x-y)_+^{2j_1+2}}{((x-y)^2)^{2j_1+4}},\label{ZZNorm}\\
\langle [O_{j_1j_1}^{ss,\ 20}]_{43}^{21}(x)\overline{[O_{j_1j_1}^{ss,\ 20}]_{43}^{21}}(y)\rangle\vert_{ _{g=0}}=\delta_{j_1,j_2}\mathcal{N}^2(N_c^2-1)\sigma_{j_1}^2\Gamma(2j_1+3)3^22^{2j_1+5}\frac{(x-y)_+^{2j_1+2}}{((x-y)^2)^{2j_1+4}}.
\end{gather}

\subsection{ Calculation at $g^2$ order}

Now we will compute this 3-point correlator  at  \(g^2 \) order. The general form  of  individual terms (coming from the expansion of Gegenbauer polynomials)  contributing to the correlator is as follows :

\begin{gather}
\langle \Tr[X,Z]^2(\gamma)\Tr\left(D^{m_{1}}_{\alpha_1}X(\alpha_1)D^{m_{2}}_{\alpha_2}X(\alpha_2)\right)
\Tr\left(D^{k_{1}}_{\beta_1}Z(\beta_1)D^{k_{2}}_{\beta_2}Z(\beta_2)\right)\rangle,
\end{gather}
where the  \(m_1,m_2,k_1,k_2\) are some integer powers.

In this case, we use the point-splitting again. Formally, it breaks gauge-invariance and we should insert little Wilson lines between separated points inside the operator. But one can easily check that such Wilson lines do not contribute to the correlator, and can be omitted.  All diagrams, which have two free propagators between \(Z(X)\) and \(\bar{Z}(\bar{X})\),  disappear after the action of differential operators, similarly to the \(g^0\) order (\ref{0inTreeL}). Thus only three types of diagrams contribute. The first one, depicted in Fig.\ref{ris:2gl-s-s V},  has two scalar-scalar-gluon vertices, and it looks like a \(g^0\) order diagram with one gluon line connecting two scalar propagators. The second diagram, depicted in Fig.\ref{ris:1gl-s-s V}, has one scalar-scalar-gluon vertex  connecting a scalar propagator with a gauge field \(A_{+}\) from covariant derivative in twist-2 operators. And the third diagram , of a type of the one depicted in Fig.\ref{ris:4Scalar}, includes the 4-scalar vertex. Let us compute  each of these contributions.

\subsubsection{Two scalar-scalar-gluon vertices}

The structure of scalar-scalar-gluon vertex can be read off  from the  term \(2\Tr D_{\mu}XD^{\mu}\bar{X}\) of the N=4 SYM Lagrangian. The relevant terms from this vertex are

\begin{gather}
-2g \Tr([A_{\mu},X]\partial^{\mu}\bar{X}+[A_{\mu},\bar{X}]\partial^{\mu}X)=\notag\\
=-2g\Tr((\partial^\mu\bar{X}A_{\mu}X-\bar{X}A_{\mu}\partial^{\mu}X)+(\partial^\mu XA_{\mu}\bar{X}-XA_{\mu}\partial^{\mu}\bar{X}))\,.\label{ssg}
\end{gather}

\begin{figure}[h]
\begin{minipage}[h]{0.49\linewidth}
\center{\includegraphics[scale=0.2]{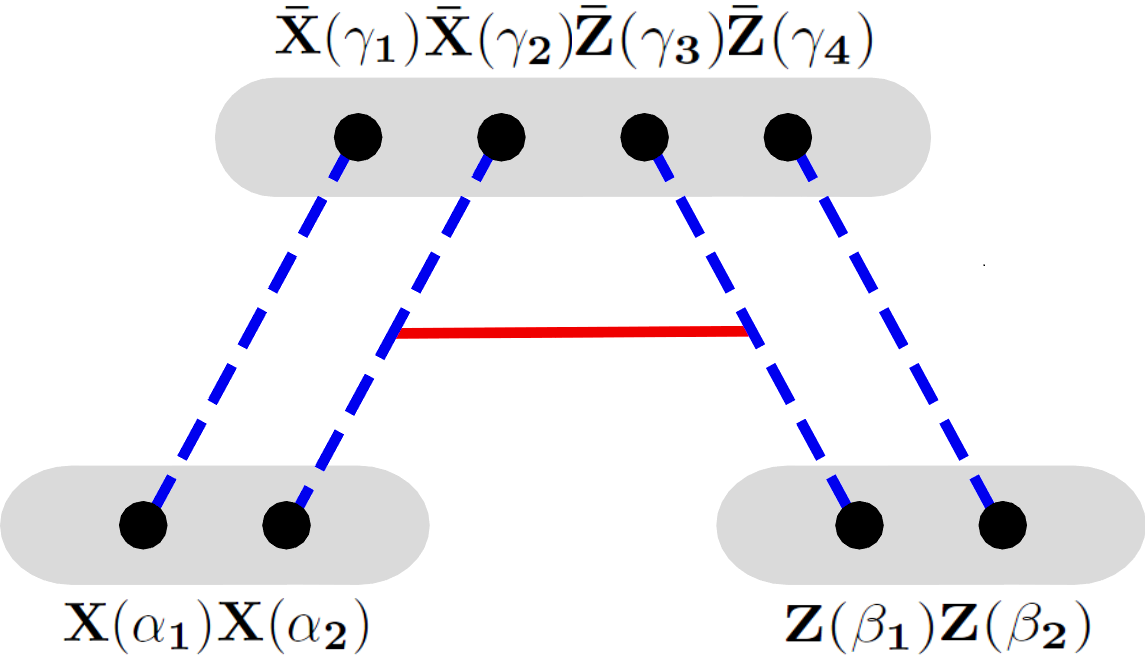}\\ (a)}
\end{minipage}
\hfill
\begin{minipage}[h]{0.49\linewidth}
\center{\includegraphics[scale=0.23]{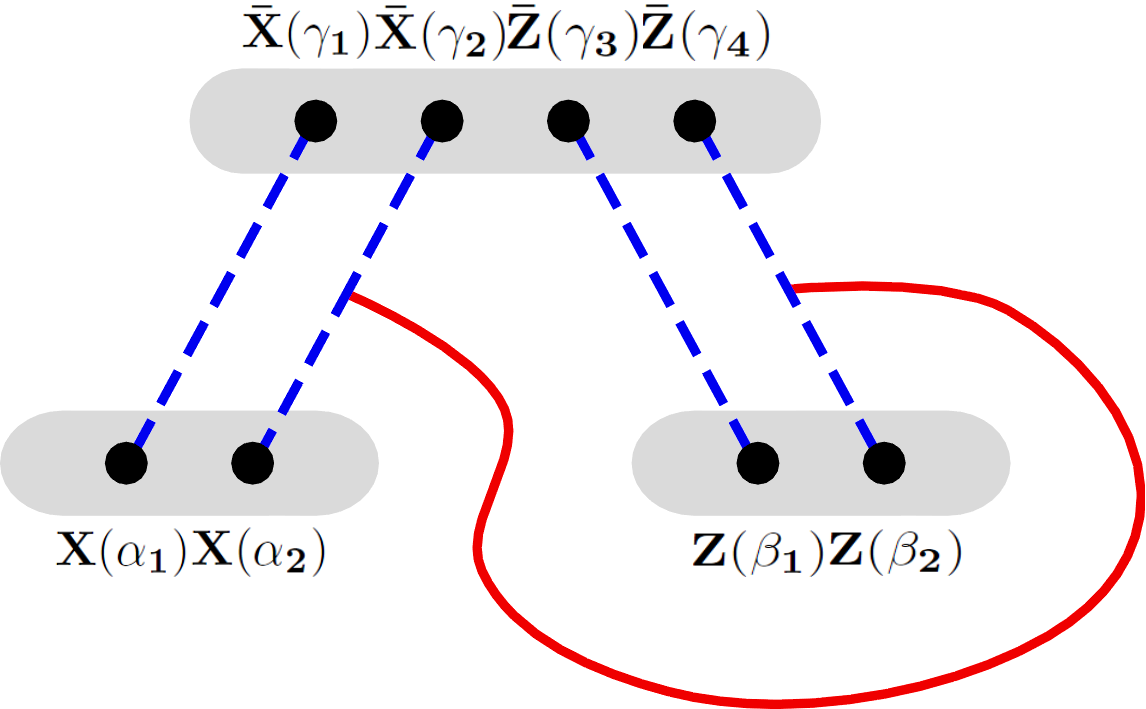} \\ (b)}
\end{minipage}
\caption{Diagrams with two scalar-scalar-gluon vertex.}
\label{ris:2gl-s-s V}
\end{figure}

Two relevant vertices, when being put down from the action in the functional integral, give
\begin{gather}
W_{ssgss}=-4g^2\int\int d^4ud^4v\Tr((\partial^\mu\bar{X}A_{\mu}X-\bar{X}A_{\mu}\partial^{\mu}X)+(\partial^\mu XA_{\mu}\bar{X}-XA_{\mu}\partial^{\mu}\bar{X}))(u)\cdot\notag\\
\cdot \Tr((\partial^\nu\bar{Z}A_{\nu}Z-\bar{Z}A_{\nu}\partial^{\nu}Z)+(\partial^\nu ZA_{\nu}\bar{Z}-ZA_{\nu}\partial^{\nu}\bar{Z}))(v).\label{ssgss}
\end{gather}
Thus for the term \(\Tr(\bar{X}^2\bar{Z}^2)\) in Konishi operator we should calculate

\begin{gather}
\langle \Tr(X(\alpha_1)X(\alpha_2))\Tr(Z(\beta_1)Z(\beta_2))\Tr(\bar{X}(\gamma_1)\bar{X}(\gamma_2)\bar{Z}(\gamma_3)\bar{Z}(\gamma_4))W_{ssgss}\rangle.
\end{gather}
Let us consider the diagram depicted on Fig.\ref{ris:2gl-s-s V}.a. From (\ref{ssgss}) we get four terms of the type \(\Tr[\partial^\mu\bar{X}A_{\mu}X-\bar{X}A_{\mu}\partial^{\mu}X]\Tr[(\partial^\nu\bar{Z}A_{\nu}Z-\bar{Z}A_{\nu}\partial^{\nu}Z)]\) and the corresponding expressions read as
follows\footnote{The symbol \(\left(\begin{smallmatrix}\alpha_1 \leftrightarrow \alpha_2,\\ \beta_1\leftrightarrow \beta_2
\end{smallmatrix}\right)\) denotes all the terms obtained from the previous one (including the sign) by all possible permutations, giving 3 extra terms  }
\begin{gather}
(-4g^2)(-2N_c^2+6-\frac{4}{N_c^2})\frac{\mathcal{N}^7}{|\gamma_1-\alpha_1|^2|\gamma_4-\beta_2|^2}V(\gamma_2,\gamma_3,\alpha_2,\beta_1)+
\left(\begin{smallmatrix}\alpha_1 \leftrightarrow \alpha_2,\\ \beta_1\leftrightarrow \beta_2
\end{smallmatrix}\right) ,\label{Fig1a1}\\
-(-4g^2)(N_c^4-5N_c^2+8-\frac{4}{N_c^2})\frac{\mathcal{N}^7}{|\gamma_1-\alpha_1|^2|\gamma_4-\beta_2|^2}V(\gamma_2,\gamma_3,\alpha_2,\beta_1)+\left(\begin{smallmatrix}\alpha_1 \leftrightarrow \alpha_2,\\ \beta_1\leftrightarrow \beta_2
\end{smallmatrix}\right),\\
-(-4g^2)(4-\frac{4}{N_c^2})\frac{\mathcal{N}^7}{|\gamma_1-\alpha_1|^2|\gamma_4-\beta_2|^2}V(\gamma_2,\gamma_3,\alpha_2,\beta_1)+\left(\begin{smallmatrix}\alpha_1 \leftrightarrow \alpha_2,\\ \beta_1\leftrightarrow \beta_2
\end{smallmatrix}\right),\\
(-4g^2)(-2N_c^2+6-\frac{4}{N_c^2})\frac{\mathcal{N}^7}{|\gamma_1-\alpha_1|^2|\gamma_4-\beta_2|^2}V(\gamma_2,\gamma_3,\alpha_2,\beta_1)+\left(\begin{smallmatrix}\alpha_1 \leftrightarrow \alpha_2,\\ \beta_1\leftrightarrow \beta_2
\end{smallmatrix}\right).\label{Fig1a4}\\
\end{gather}
where we have introduced the function
\begin{gather}
V(x_1,x_2,x_3,x_4)=\notag\\
=(\partial_3-\partial_1)(\partial_2-\partial_4)\int\int d^4u d^4v\frac{1}{|x_1-u|^2|x_3-u|^2|u-v|^2|x_2-v|^2|x_4-v|^2}.
\end{gather}

\vskip.5cm
\vskip1cm

For Fig.\ref{ris:2gl-s-s V}.b. we get

\begin{gather}
(-4g^2)(N_c^4-5N_c^2+8-\frac{4}{N_c^2})\frac{\mathcal{N}^7}{|\gamma_1-\alpha_1|^2|\gamma_4-\beta_2|^2}V(\gamma_2,\gamma_3,\alpha_2,\beta_1)+\left(\begin{smallmatrix}\alpha_1 \leftrightarrow \alpha_2,\\ \beta_1\leftrightarrow \beta_2
\end{smallmatrix}\right),\label{Fig1b1}\\
-(-4g^2)(-2N_c^2+6-\frac{4}{N_c^2})\frac{\mathcal{N}^7}{|\gamma_1-\alpha_1|^2|\gamma_4-\beta_2|^2}V(\gamma_2,\gamma_3,\alpha_2,\beta_1)+\left(\begin{smallmatrix}\alpha_1 \leftrightarrow \alpha_2,\\ \beta_1\leftrightarrow \beta_2
\end{smallmatrix}\right),\\
-(-4g^2)(-2N_c^2+6-\frac{4}{N_c^2})\frac{\mathcal{N}^7}{|\gamma_1-\alpha_1|^2|\gamma_4-\beta_2|^2}V(\gamma_2,\gamma_3,\alpha_2,\beta_1)+\left(\begin{smallmatrix}\alpha_1 \leftrightarrow \alpha_2,\\ \beta_1\leftrightarrow \beta_2
\end{smallmatrix}\right),\\
(-4g^2)(4-\frac{4}{N_c^2})\frac{\mathcal{N}^7}{|\gamma_1-\alpha_1|^2|\gamma_4-\beta_2|^2}V(\gamma_2,\gamma_3,\alpha_2,\beta_1)+\left(\begin{smallmatrix}\alpha_1 \leftrightarrow \alpha_2,\\ \beta_1\leftrightarrow \beta_2
\end{smallmatrix}\right) \label{Fig1b4}.
\end{gather}

Each term from Fig.\ref{ris:2gl-s-s V}.a. has a partner from Fig.\ref{ris:2gl-s-s V}.b with opposite sign and a function \(V\) with different first two arguments. It turns out that the difference between them doesn't lead to nonzero contribution~\footnote{In the planar limit this phenomenon was first noticed in \cite{Okuyama:2004bd,Alday:2005nd}}. Let us show it by rewriting the function  \(V_{Euclid}(x_1,x_2,x_3,x_4)\), which is the analytic continuation of \(V(x_1,x_2,x_3,x_4)\), in the following way \cite{Beisert:2002bb} in Euclidean space:

\begin{gather}\label{V-Euclid}
V_{Euclid}(x_1,x_2,x_3,x_4)=(s-r)\frac{\phi(r,s)}{x_{13}^2x_{24}^2}+(r_1-s_1)\frac{\phi(r_1,s_1)}{x_{13}^2x_{24}^2}+\notag\\
+(r_2-s_2)\frac{\phi(r_2,s_2)}{x_{13}^2x_{24}^2}+(r_3-s_3)\frac{\phi(r_3,s_3)}{x_{13}^2x_{24}^2}+(r_4-s_4)\frac{\phi(r_4,s_4)}{x_{13}^2x_{24}^2},
\end{gather}
where
\begin{gather}
r=\frac{x_{12}^2x_{34}^2}{x_{13}^2x_{24}^2},\ \ s=\frac{x_{14}^2x_{23}^2}{x_{13}^2x_{24}^2},\ r_1=\frac{x_{34}^2}{x_{24}^2},s_1=\frac{x_{23}^2}{x_{24}^2},\notag\\
r_2=\frac{x_{34}^2}{x_{13}^2},\ s_2=\frac{x_{14}^2}{x_{13}^2}, \ r_3=\frac{x_{12}^2}{x_{24}^2}, s_3=\frac{x_{14}^2}{x_{24}^2},
r_4=\frac{x_{12}^2}{x_{13}^2},s_4=\frac{x_{23}^2}{x_{13}^2}.
\end{gather}
And the function \(\phi (r,s)=\int \limits_0^1dx \frac{-\ln(\frac{r}{s})-2\ln x}{s-x(r+s-1)+x^2r}\) \cite{Usyukina:1992jd} has a simple asymptotics when \(r\rightarrow 0, s\rightarrow 1\) (for \(x_1 \rightarrow x_2\))
\begin{gather}
\phi (r,s)=2+\log \frac{1}{r}+\cO\left((1-s)\ln r\right).
\end{gather}
In our case we should replace    \(x_1-x_2\) in (\ref{V-Euclid}) by \(\gamma_{23}=\gamma_2-\gamma_3\)  for (\ref{Fig1a1})-(\ref{Fig1a4})  and \(\gamma_{24}=\gamma_2-\gamma_4\) for (\ref{Fig1b1})-(\ref{Fig1b4}). In the limit when all points \(\gamma_i \to \gamma\) when  \(\gamma_{23}=c_{23}\epsilon ,\ \gamma_{24}=c_{24}\epsilon \) with the overall scale \(\epsilon \to 0\) and fixed \(c_{23}, c_{24}\), we observe the cancelation of \(\log\)-divergent terms  and the rest is proportional to the product of 4 propagators \(\sim\frac{\log(c_{23}/c_{24})}{|\gamma-\alpha_1|^2|\gamma-\alpha_2|^2|\gamma-\beta_1|^2|\gamma-\beta_2|^2}\). These terms disappear  when acting on them by  differential operators (acting only on \(\a_{1,2},\b_{1,2}\), and not on \(\gamma_j\)), as in the case of \(g^0\) order. For the second
term in Konishi operator \(\Tr \bar{X}\bar{Z}\bar{X}\bar{Z}\) we have the terms similar to (\ref{Fig1a1}) -(\ref{Fig1a4}), (\ref{Fig1b1})-(\ref{Fig1b4}), but they have the same color-factor \(\pm (-N_c^2+5-\frac{4}{N_c^2})\), and are
subject to   similar cancelations. Finally, the answer for the sum of diagrams with two scalar-scalar-gluon vertices is equal to zero.

\subsubsection{One scalar-scalar-gluon vertex}

\begin{figure}[h]
\begin{minipage}[h]{0.49\linewidth}
\center{\includegraphics[scale=0.2]{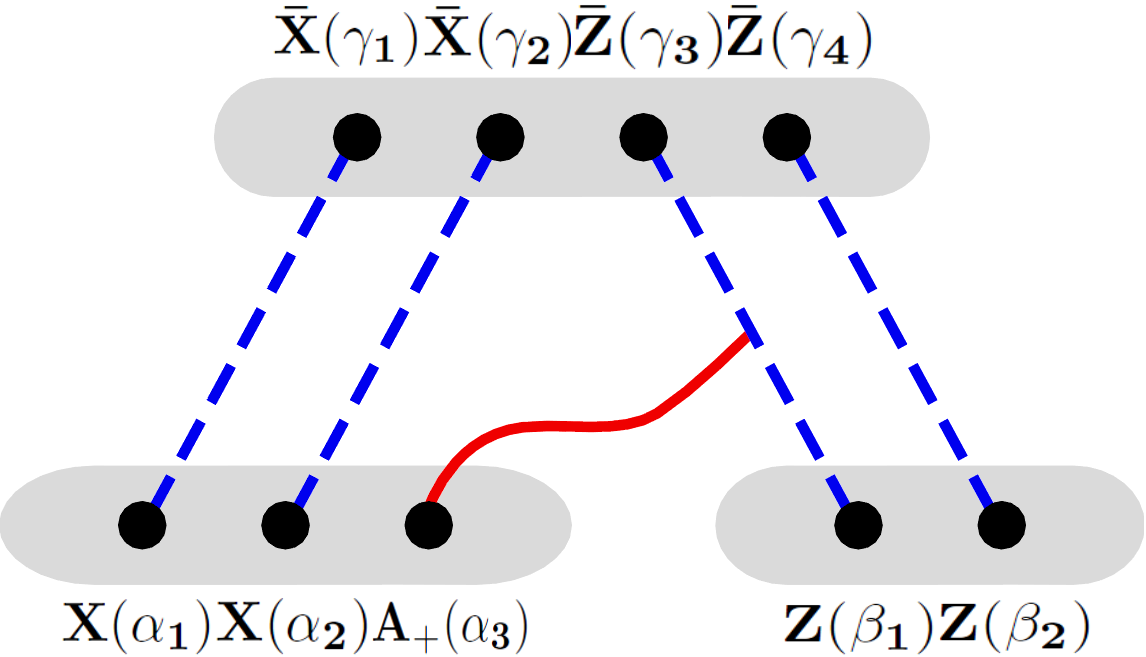} \\ (a)}
\end{minipage}
\hfill
\begin{minipage}[h]{0.49\linewidth}
\center{\includegraphics[scale=0.23]{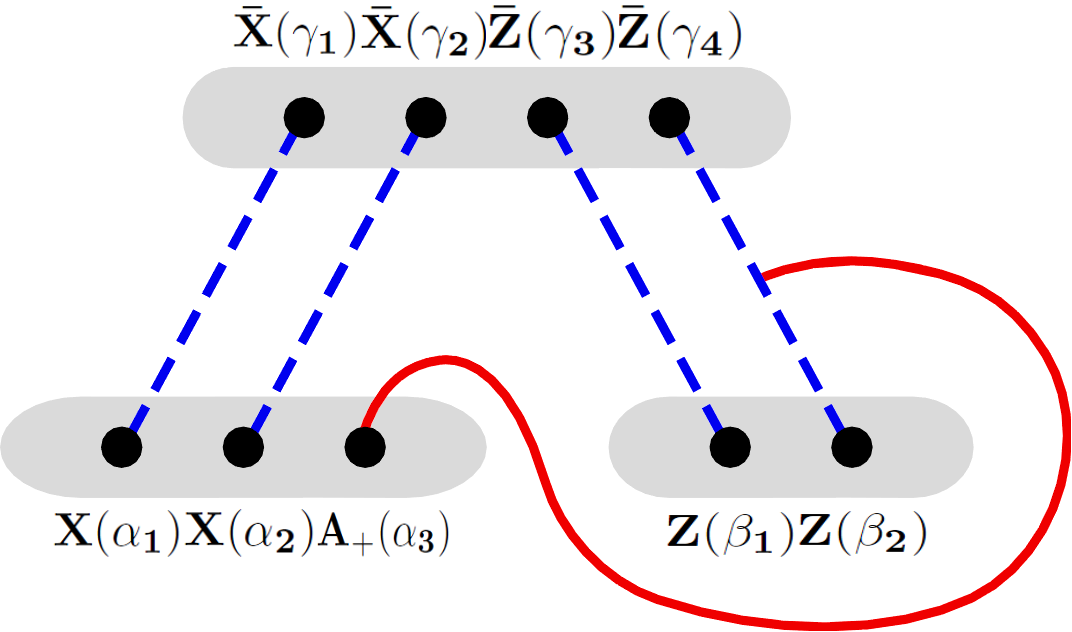} \\ (b)}
\end{minipage}
\caption{Diagrams with one scalar-scalar-gluon vertex.}
\label{ris:1gl-s-s V}
\end{figure}

\begin{figure}
\center{\includegraphics[scale=0.2]{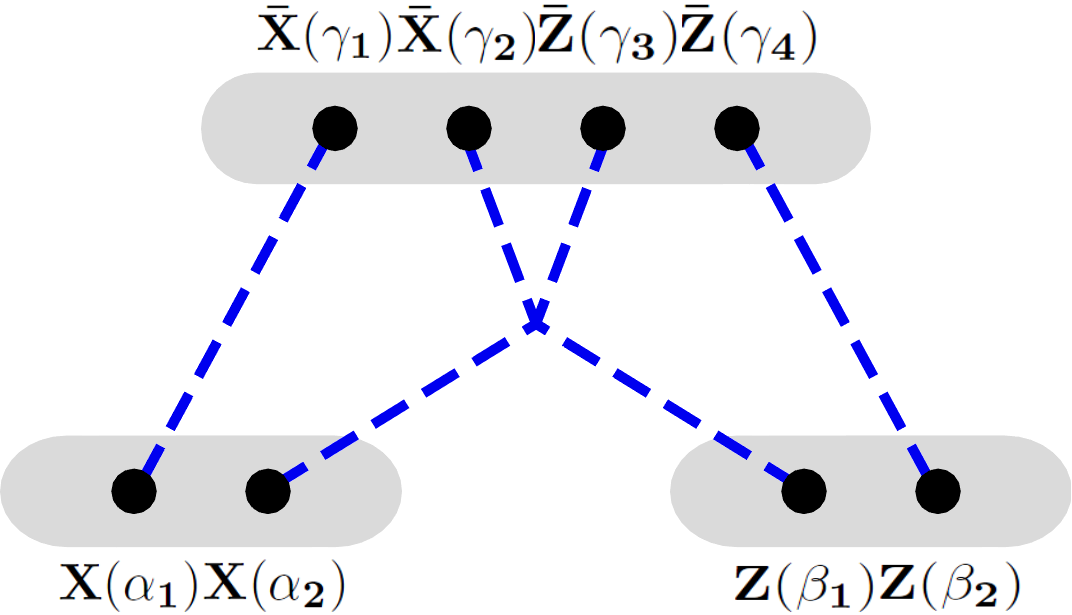} \\}
\caption{Diagram with a four-scalar vertex. }
\label{ris:4Scalar}
\end{figure}

In this case the scalar-scalar-gluon vertex connects one scalar propagator to a gluon \(A_+\) from covariant derivative, which is symbolically  depicted for \(Tr\bar{X}^2\bar{Z}^2\) in Fig.\ref{ris:1gl-s-s V}.
In this case, we need only the first two terms of expansion over \(g\) for \(D_+^nX=\partial_+^nX-ig\sum\limits_{k=1}^{n}C_n^k[\partial^{k-1}A_+,\partial^{n-k}X]\), and by factoring out the binomial coefficient and all derivatives, we get \(\Tr(XXA_+)\) inside the functional integral. As in the previous case, we get full cancellation of all terms in all orders of \(N_c\), both for \(\Tr\bar{X}^2\bar{Z}^2\) and \(\Tr\bar{X}\bar{Z}\bar{X}\bar{Z}\) terms of Konishi. For the details see appendix \ref{App1loop}.

\subsubsection{4-scalar vertex and the full $g^2$ order result}

Let us write explicitly the relevant term in the Lagrangian responsible for the 4-scalar interaction (between \(X\) and \(Z\) components):
\begin{gather}
\frac{1}{8}g^2\Tr[\phi^{AB},\phi^{CD}][\bar{\phi}^{AB},\bar{\phi}^{CD}]=\notag\\
2g^2\Tr(2Z\bar{X}\bar{Z}X+2\bar{X}ZX\bar{Z}-Z\bar{X}X\bar{Z}-\bar{X}Z\bar{Z}X
-ZX\bar{X}\bar{Z}-XZ\bar{Z}\bar{X}). \label{4-s.v.}
\end{gather}
Carrying out explicit calculation (see Appendix \ref{App1loop} for  details)  we get the following final expression for this 3-point correlator in the first non-vanishing, Born contribution:

\begin{gather}
K_{j_1j_2}(\a,\b,\gamma)\equiv\langle [O_{j_1j_1}^{ss,\ 20}]_{41}^{32}(\alpha)[O_{j_2j_2}^{ss,\ 20}]_{43}^{21}(\beta)\Tr[X,Z]^2(\gamma) \rangle=\notag\\
=-\sigma_{j_1}\sigma_{j_2}3^32^6\pi^2\mathcal{N}^6g^2(N_c^4-N_c^2)\mathcal{G}^{\frac{1}{2};g=0}_{j_1+1,\alpha_1,\alpha_2}\mathcal{G}^{\frac{1}{2};g=0}_{j_2+1,\beta_1,\beta_2}\Psi|_{\begin{smallmatrix}\alpha_{1,2} =\alpha,\\ \beta_{1,2}=\beta\end{smallmatrix}},\label{1loopAnswer}
\end{gather}
where \(\Psi\) is defined as
\begin{gather}
\Psi=\frac{\log\frac{|\alpha_2-\gamma|^2|\beta_2-\gamma|^2}{|\alpha_2-\beta_2|^2|\epsilon|^2}}{|\alpha_1-\gamma|^2|\alpha_2-\gamma|^2|\beta_1-\gamma|^2|\beta_2-\gamma|^2}+
\left(\begin{smallmatrix}\alpha_1 \leftrightarrow \alpha_2,\\ \beta_1\leftrightarrow \beta_2\end{smallmatrix}\right).\label{OmegaExpr}
\end{gather}
This expression looks somewhat similar to the \(g^0\) order  expressions for three twist-2 correlators. It is given by the action of differential operators  defining the form of the primary operators by the action on a simple function. The latter is just a product of free propagators in the \(g^0\) order calculation, and in the current case it is expressed by the function  \(\Psi\) containing an extra \(\log\). Moreover,  the expression  \eqref{1loopAnswer} doesn't have any \(\log\) terms as it should be in our case when the three-point correlator is zero in the \(g^0\) order. All logarithms can  come only from the series expansion of three-point correlator in powers of anomalous dimensions, but this expansion starts at least from \(g^2\). Technically, one can check this cancellation in (\ref{1loopAnswer}) acting directly by differential operators and summing up all terms which have a logarithm. Due to the cancelation of logarithms by action of any derivative, one should keep only those terms which do not act on logarithm. These terms have the form \(\log(...)\mathcal{K*}\frac{1}{|\alpha_1-\gamma|^2|\alpha_2-\gamma|^2|\beta_1-\gamma|^2|\beta_2-\gamma|^2}\)  and  their sum turns out to be equal  zero, by the same reasons as at the \(g^0\) order.

Generically, our correlator contains many terms which correspond to different tensor structures (appendix \ref{AppCorWithSpins}) as it was in the case of  three twist-2 operators. If we restrict the coordinates of the points to the  2-d plane \((n_+,n_-)\) all tensor structures factorize again into a single one  (\ref{T2T2Kspinstr})
\begin{equation}
K_{\natural j_1j_2}(\a,\b,\gamma)=\frac{B_{\natural j_1j_2}}{(\alpha-\gamma)_+^2(\beta-\gamma)_+^2(\alpha-\gamma)_-^{2+j_1-j_2}
(\beta-\gamma)_-^{2+j_2-j_1}(\alpha-\beta)_-6^{j_1+j_2+2}}.
\end{equation}
Now, if we send \(\gamma_-\) to infinity its asymptotic form reads as follows

\begin{gather}\label{3-pT2KT2}
K_{\natural j_1j_2}(\a,\b,\gamma)\simeq\frac{B_{\natural j_1j_2}}{(\alpha-\gamma)_+^2(\beta-\gamma)_+^2(\gamma)_-^{4}
(\alpha-\beta)_-^{j_1+j_2+2}}.
\end{gather}
On the other hand, this asymptotics can be obtained directly from (\ref{1loopAnswer}). Indeed, the leading power in \(\gamma_-\) corresponds to the maximal powers of \((\alpha\beta)_-\) in denominator and  appear when all \(j_1+j_2+2\) derivatives acts on the \(\log\). It gives us

\begin{gather}
K_{\natural j_1j_2}(\a,\b,\gamma)=\sigma_{j_1}\sigma_{j_2}3^32^4\pi^2\mathcal{N}^6(N_c^4-N_c^2)g^2\frac{i^{j_1+j_2}\Gamma(j_1+j_2+2)}{(\alpha-\gamma)_+^2(\beta-\gamma)_+^2(\gamma)_-^{4}
(\alpha-\beta)_-^{j_1+j_2+2}}.
\end{gather}
Comparing it with \eqref{3-pT2KT2} we conclude that
\begin{gather}
B_{\natural j_1j_2}=g^2i^{j_1+j_2}\sigma_{j_1}\sigma_{j_2}3^32^4\pi^2\mathcal{N}^6(N_c^4-N_c^2)\Gamma(j_1+j_2+2).
\end{gather}
Normalizing this 3-point function by the corresponding two-point functions (\ref{KonishiNorm}), (\ref{ZZNorm}) we finally obtain for the normalized structure constant \(C_{\natural j_1j_2}\) of two twist-2 operators and Konishi operator:

\begin{gather}
C_{\natural j_1j_2}=g^2\sigma_{j_1}\sigma_{j_2}3^{\frac{1}{2}}2^{-4}\pi^{-2}\frac{N_c}{\sqrt{N_c^2-1}}i^{j_1+j_2}\frac{\Gamma(j_1+j_2+2)}{\left(\Gamma(2j_1+3)\Gamma(2j_2+3)\right)^{\frac{1}{2}}}\,\,.
\end{gather}

\section{Conclusions}

In this paper, we have explicitly calculated in the leading, Born approximation two types  of 3-point correlators of the \(SU(N_c)\) conformal N=4 SYM theory at finite \(N_c\). One of them involves three twist-2 operators with arbitrary spins and another - two twist-2 operators   and  one Konishi operator (namely, its component built from scalar fields).   The related structure constants show a very rich structure of the operator product expansion of the theory, even at this approximation.   The \(g^0\) order  result for 3 twist-2 operators for    the correlators with arbitrary spins are given by the action of certain differential operators (built from Gegenbauer polynomials) on a standard correlator with the  lowest spins. It is interesting that the second correlator, involving Konishi operator and calculated at the leading \(g^2\) approximation,  shows the same pattern. This and some other features of our results may provide some insight for the future generalizations to higher orders in the YM coupling and to other operators. Some of the general structures of the OPE at all orders were  conjectured from the  study of correlators of \(SU(2)\) sector involving only scalar fields in  \cite{Gromov:2012uv,Serban:2012dr}. We hope that our attempt to extend calculations to the operators containing covariant derivatives may shed some light on the properties of OPE for the full set of operator of the theory.   Of course, if we want integrability to help us on this way we should go to the planar limit. On the other hand, it would be curious to compute, using these structure constants, the multi-point correlators of twist-2 operators in the Born approximation and investigate on this "simple" example some general properties of the full OPE. Another interesting problem to solve is to include higher twist operators, at least at the leading \(g^0\) order.

It would be also interesting to compare our results to the strong coupling calculations of similar correlators (corresponding to the  GKP
 states on the string side) performed in \cite{Kazama:2012is}. The details of this comparison contain subtle questions of normalization of operators which have to be elucidated prior to such comparison.

\section*{Acknowledgements}
We thank     N.Gromov,  I.Kostov, R.Janik,  D.Serban,  A.Sever, P.Vieira, and K.Zarembo  for interesting discussions.   The  comments  of N.Gromov,     D.Serban,  P.Vieira, and K.Zarembo   to the manuscript was very useful.  We are especially grateful to G.Korchemsky  who  taught us a lot of very useful facts at various stages of this project and carefully read the manuscript.
Our work  was also partly supported by the ANR grants StrongInt (BLANC-SIMI-
4-2011)  and by the ESF grants HOLOGRAV-09-RNP-
092 and ITGP.
 We are thankful to  the Israel Institute for Advanced Studies in Jerusalem and to  {\it\ Institut f¨\"ur Mathematik und Institut f¨\"ur Physik, Humboldt-Universit¨at zu Berlin} for hospitality during the initial stage
of this work.  V.K. thanks the Alexander von Humboldt foundation for the support.

\section*{Appendices}

\appendix

\section{Notations}\label{AppNotations}

In this section we set our notations.
The lagrangian of N=4 SYM with the \(SU(N_c)\) gauge group has the following form:

\begin{gather}
\mathfrak{L}=\Tr\left\{-\frac{1}{2}F_{\mu\nu}F^{\mu\nu}+\frac{1}{2}(D_\mu\phi^{AB})(D^\mu\bar{\phi}_{AB}) +\frac{1}{8}g^2[\phi^{AB},\phi^{CD}][\bar{\phi}_{AB},\bar{\phi}_{CD}]+\right. \notag\\
\left.+2i\bar{\lambda}_{\dot{\alpha}A}\sigma_{\mu}^{\dot{\alpha}\beta}D^\mu\lambda_\beta^A-\sqrt{2}g\lambda^{\alpha A}[\bar{\phi}_{AB},\lambda^B_\alpha]+\sqrt{2}g\bar{\lambda}_{\dot{\alpha}A}[\phi^{AB},\bar{\lambda}_B^{\dot{\alpha}}]\right\},
\end{gather}
where field strength \(F_{\mu\nu}=\partial_\mu A_\nu-\partial_\nu A_\mu-ig[A_\mu,A_\nu]\) and covariant derivative \(D_\mu=\partial_\mu -ig[A_\mu,...]\).Notice that we work with Minkowski signature \((+,-,-,-)\) and all fields are taken in the adjoint representation of \(SU(N_c)\). \(SO(6)\)-multiplet with scalars \(\phi^a, a\in\{1\div6\}\) can be grouped into the antisymmetric tensor \(\phi^{AB}\),\(A,B\in\{1\div4\}\):
\begin{equation}
\phi^{AB}=\frac{1}{\sqrt{2}}\Sigma^{a AB}\phi^a, \ \ \ \ \bar{\phi}_{AB}=\frac{1}{\sqrt{2}}\bar{\Sigma}^a_{AB}\phi^a=(\phi^{AB})^*,
\end{equation}
using Dirac matrices in 6-d Euclidian space:
\begin{gather*}
\Sigma^{a AB}=(\eta_{1AB},\eta_{2AB},\eta_{3AB},i\bar{\eta}_{1AB},i\bar{\eta}_{2AB},i\bar{\eta}_{3AB}),\\
\bar{\Sigma}^a_{AB}=(\eta_{1AB},\eta_{2AB},\eta_{3AB},-i\bar{\eta}_{1AB},-i\bar{\eta}_{2AB},-i\bar{\eta}_{3AB}),
\end{gather*}
and 't Hooft symbols:
\begin{gather*}
\eta_{iAB}=\epsilon_{iAB}+\delta_{iA}\delta_{4B}-\delta_{iB}\delta_{4A},\\
\bar{\eta}_{iAB}=\epsilon_{iAB}-\delta_{iA}\delta_{4B}+\delta_{iB}\delta_{4A},
\end{gather*}

\begin{equation}
\eta_1=\begin{pmatrix}
0 & 0 & 0 & 1\\
0& 0 & 1 & 0\\
0& -1 & 0 & 0\\
-1 & 0&0&0\\
\end{pmatrix},
\eta_2=\begin{pmatrix}
0 & 0 & -1 & 0\\
0& 0 & 0 & 1\\
1& 0 & 0 & 0\\
0 & -1& 0& 0\\
\end{pmatrix},
\eta_3=\begin{pmatrix}
0 & 1 & 0 & 0\\
-1& 0 & 0 & 0\\
0& 0 & 0 & 1\\
0 & 0& -1& 0\\
\end{pmatrix},
\end{equation}

\begin{equation}
i\bar{\eta}_1=\begin{pmatrix}
0 & 0 & 0 & -i\\
0& 0 & i & 0\\
0& -i & 0 & 0\\
i & 0&0&0\\
\end{pmatrix},
i\bar{\eta}_2=\begin{pmatrix}
0 & 0 & -i & 0\\
0& 0 & 1 & -i\\
i& 0 & 0 & 0\\
0 & i& 0& 0\\
\end{pmatrix},
i\bar{\eta}_3=\begin{pmatrix}
0 & i & 0 & 0\\
-i& 0 & 0 & 0\\
0& 0 & 0 & -i\\
0 & 0& i& 0\\
\end{pmatrix}.
\end{equation}
Explicit formula for scalars reads as follows

\begin{gather*}
[\phi^{AB}]=\frac{1}{\sqrt{2}}(\phi^1 \eta_{1AB}+\phi^2\eta_{2AB}+\phi^3\eta_{3AB}+\phi^4i\bar{\eta}_{1AB}+\phi^5i\bar{\eta}_{2AB}+\phi^6i\bar{\eta}_{3AB})=\\
=\frac{1}{\sqrt{2}}\begin{pmatrix}
0 & \phi^3+i\phi^6 & -\phi^2-i\phi^5 & \phi^1-i\phi^4\\
-\phi^3-i\phi^6& 0 & \phi^1+i\phi^4 & \phi^2-i\phi^5\\
\phi^2+i\phi^5& -\phi^1-i\phi^4 & 0 & \phi^3-i\phi^6\\
-\phi^1+i\phi^4 & -\phi^2+i\phi^5& -\phi^3+i\phi^6& 0\\
\end{pmatrix}
=\begin{pmatrix}
0 & Z & -Y & \bar{X}\\
-Z& 0 & X & \bar{Y}\\
Y& -X & 0 & \bar{Z}\\
-\bar{X} & -\bar{Y}& -\bar{Z}& 0\\
\end{pmatrix}.
\end{gather*}
Fermions are realized as a two-component Weyl spinors \(\lambda_\alpha^A\) with conjugated \(\bar{\lambda}_{\dot{\alpha}A}\). Spinor index \(\alpha\in\{1,2\}\)and \(A\in\{1\div4\}\) is a \(SU(4)\) index. Due to supersymmetry one can fix just the propagator of scalars and get the normalization for fermions and gauge fields acting by supercharges. In this article we set the normalization for free propagators as follows:

\begin{gather}
\langle Z(x)^a_b \bar{Z}(y)^c_d\rangle=\mathcal{N}(\delta ^a_d\delta ^c_b-\frac{1}{N_c}\delta ^a_b\delta ^c_d)\frac{1}{(x-y)^2},\ \ \text{and the same for}\  X \text{and}\  Y,\\
\langle \lambda_{\alpha}^A(x)^a_b \bar{\lambda}_{\dot{\beta} B}(y)^c_d \rangle=i\mathcal{N}\delta^A_B(\delta ^a_d\delta ^c_b-\frac{1}{N_c}\delta ^a_b\delta ^c_d)\bar{\sigma}^\mu_{\alpha \dot{\beta}}\frac{\partial}{\partial x^\mu}\frac{1}{(x-y)^2},\\
\langle A_\mu(x)^a_b A_\nu(y)^c_d \rangle =-\mathcal{N}(\delta ^a_d\delta ^c_b-\frac{1}{N_c}\delta ^a_b\delta ^c_d)\frac{g_{\mu \nu}}{(x-y)^2}.
\end{gather}
where \(\mathcal{N}=-\frac{1}{8\pi^2}\), \(\{\sigma^\mu\}=\{1,\mathbf{\sigma}\}\) and \(\{\bar{\sigma}^\mu\}=\{1,-\mathbf{\sigma}\}\) with ordinary Pauli matrices \(\mathbf{\sigma}\).
Throughout the text we use the basis \(\{n_+,n_-,e_{1\bot},e_{2\bot}\}\) with two  light-like vectors \(n_+^\mu=\{\frac{1}{\sqrt{2}},0,0,\frac{1}{\sqrt{2}}\},\ \ n_-^\mu=\{\frac{1}{\sqrt{2}},0,0,-\frac{1}{\sqrt{2}}\}\) normalized as \((n_- n_+)=1\) and two orthogonal vectors \(e_{1\bot},e_{2\bot}\) , which span 2-d plane \(\{\bot\}\) orthogonal to \(\{n_+,n_-\}\). The vector \(x\) reads in this basis  as      \(x=x_- n_+ +x_+n_-+x_\bot\).

\par\medskip
\subsection{Field content of twist-2 operators}
All twist-2 operators, which were discussed in this paper, are constructed from the set of elementary fields \(X=\{F^{+\mu}_{\ \ \bot}, \lambda^A_{+\alpha},\bar{\lambda}^{\dot{\alpha}}_{+A},\phi^{AB}\}.\)  Twist 2 is the minimal possible twist (defined as dimension minus spin). Gluon field \(F^{+\mu}_{\ \ \bot}\) is obtained by projection of one of the indices of the field strength tensor \(F^{\mu\nu}\) on \(n^+\) direction where as  the second index is automatically restricted to the transverse plane with the metric \(g^\bot_{\mu\nu}=g_{\mu\nu}-n_{+\mu} n_{-\nu}-n_{+\nu} n_{-\mu}\). Weyl spinors \(\lambda_{+\alpha}\) and \(\bar{\lambda}_+^{\dot{\alpha}}\) correspond to the states with definite helicity \(1,-1\), respectively and they are parameterized as \(\lambda_{+\alpha}=\frac{1}{2}\bar{\sigma}^-_{\alpha\dot{\beta}}\sigma^{+\dot{\beta}\gamma}\lambda_\gamma\) and \(\bar{\lambda}_+^{\dot{\alpha}}=\frac{1}{2}\sigma^{-\dot{\alpha}\beta}\bar{\sigma}^+_{\beta \dot{\gamma}}\bar{\lambda}^{\dot{\gamma}}\).

\section{Twist-2 operators at $g^0$ order  and Gegenbauer polynomials}\label{AppGeg}

 Explicit formula for conformal twist-2 operators can be obtained from the fact that they are primaries of \(SL(2,R).\) It was obtained in \cite{Makeenko:1980bh,Ohrndorf:1981qv} and can be expressed in terms of  the Jacobi polynomials \(P_{n}^{(2j_1-1,2j_2-1)}(z) \)

\begin{equation}
O^{j_1,j_2}_n(x)=X_{j_1}(x)i^n(\overleftarrow{D_+}+\overrightarrow{D_+})^n P_{n}^{(2j_1-1,2j_2-1)}\left(\frac{\overleftarrow{D_+}-
\overrightarrow{D_+}}{\overleftarrow{D_+}+\overrightarrow{D_+}}\right)X_{j_2}(x), \label{ConfPr}
\end{equation}     where  \(j_1,j_2\) are the conformal spins  and the derivatives \(\overleftarrow{D_+},\overrightarrow{D_+}\) act in light-like direction \(n_+\) on the arguments of the functions \(X_{j_1}(x)\) and \(X_{j_2}(x)\), respectively.

Gegenbauer polynomials are a particular case of Jacobi polynomials
\begin{gather}
C_n^\a(z) =\frac{\Gamma(n+2\alpha)\Gamma(1/2+\alpha)}{\Gamma(2\alpha \Gamma(n+\alpha+1/2))}P_n^{\a-\frac{1}{2},\a-\frac{1}{2}}(z),
\end{gather}
or, explicitly:

\begin{equation}\label{Gegenbauer_def}
C_n^{\alpha}(z)=\sum\limits_{k=0}^{[\frac{n}{2}]}\frac{(-1)^k(\alpha)_{n-k}(2z)^{n-2k}}{k!(n-2k)!},\ \ (\alpha)_m=\frac{\Gamma(m+\alpha)}{\Gamma(\alpha)},\end{equation}
with the orthonormality property
\begin{equation}\label{orthonorm_Gegenbauer}
\int\limits_{-1}^1(1-z^2)^{\alpha-\frac{1}{2}}C_{m}^{\alpha}(z)C_{n}^{\alpha}(z)dz=\delta_{m,n}\frac{\pi 2^{1-2\alpha}\Gamma(n+2\alpha)}{n!(n+\alpha)\Gamma(\alpha)^2}\,\,.
\end{equation}
In this paper we have used the following formulae:
\begin{gather}
(b+a)^nC^{\frac{1}{2}}_n(\frac{b-a}{b+a})=\sum\limits_{k=0}^n(-a)^{n-k}b^k{\binom{n}{k}}^2,\label{G12prod}\\
(b+a)^nC^{\frac{3}{2}}_n(\frac{b-a}{b+a})=(n+1)C_n^{\frac{3}{2}}(1)\sum\limits_{k=0}^n\frac{a^{n-k}b^k(-1)^{n-k}{\binom{n}{k}}n!}{(k+1)!(n-k+1)!}=\notag\\
=\frac{n+1}{2}\sum\limits_{k=0}^n (-a)^{n-k}b^k \binom{n}{k}\binom{n+2}{k+1},\label{G32prod}\\
(b+a)^nC^{\frac{5}{2}}_n(\frac{b-a}{b+a})=2(n+1)(n+2)C_n^{\frac{5}{2}}(1)\sum\limits_{k=0}^n\frac{a^{n-k}b^k(-1)^{n-k}{\binom{n}{k}}n!}{(k+2)!(n-k+2)!}=\notag\\
=\frac{2(n+1)(n+2)}{4!}\sum\limits_{k=0}^n (-a)^{n-k}b^k\binom{n}{k}\binom{n+4}{k+2}\label{G52prod}
\end{gather}
which can be proved from the explicit formula \ref{Gegenbauer_def} for definition of Gegenbauer polynomials.
\par\medskip
To calculate the 2-point correlator  at \(g^0 \) order   we used integral representations. Say, for gluons it looks as follows
\begin{gather}
\label{mains}
\langle \mathcal{O}^{gg}_{j_1}(x) \mathcal{O}^{gg}_{j_2}(y) \rangle= \notag \\
=\sigma_{j_1}\sigma_{j_2}\mathcal{N}^2(N^2-1)\ \mathcal{G}^{\frac{5}{2}}_{j_1-1,x_1,x_2}\enskip\mathcal{G}^{\frac{5}{2}}_{j_2-1,y_1,y_2}\int \limits_0^{\infty}\int \limits_0^{\infty} ds_1 ds_2 \frac{s_1^2s_2^2 e^{-s_1(x_1-y_1)_--s_2(x_2-y_2)_+}}{4(x_1-y_1)_+(x_2-y_2)_+}.
\end{gather}
Using the evenness of Gegenbauer polynomials, we can rewrite (\ref{mains}) as
\begin{eqnarray}
&&\langle \mathcal{O}^{gg}_{j_1}(x) \mathcal{O}^{gg}_{j_2}(y)\rangle=\sigma_{j_1}\sigma_{j_2}\mathcal{N}^2(N^2-1)i^{j_1+j_2-2}\times\notag\\
&&\times\int \limits_0^{\infty} \int \limits_0^{\infty} (s_2+s_1)^{j_1+j_2-2}C^{\frac{5}{2}}_{j_1-1}(\frac{s_2-s_1}{s_2+s_1})C^{\frac{5}{2}}_{j_2-1}(\frac{s_2-s_1}{s_2+s_1})\frac{e^{-(s_1+s_2)(x-y)_-}}{(x-y)_+^2}\frac{s^2_1s^2_2}{4}=
\nn\\
&&=-\frac{\sigma_{j_1}\sigma_{j_2}\mathcal{N}^2(N^2-1)}{4i^{j_1+j_2}(x-y)_+^2}\int \limits_0^{\infty}ds\int\limits_0^1 d\alpha s s^{j_1+j_2+2}C^{\frac{5}{2}}_{j_1-1}(1-2\alpha)C^{\frac{5}{2}}_{j_2-1}(1-2\alpha)\alpha^2 (1-\alpha)^2e^{-s(x-y)_-}=\nn\\
&&=\delta_{j_1,j_2}\mathcal{N}^2\frac{\sigma_{j_1}^2(N^2-1)i^{2j_1-2}}{4}\frac{\Gamma(2j_1+4)}{(x-y)_+^2(x-y)_-^{2j_1+4}}\frac{1}{2^5}\frac{\pi 2^{-4}\Gamma(j_1+4)}{\Gamma(j_1)(j_1-1+\frac{5}{2})\Gamma^2(\frac{5}{2})}=\nn\\
&&=\delta_{j_1,j_2}\mathcal{N}^2\frac{\sigma_{j_1}^2\Gamma(2j_1+4)\Gamma(j_1+4)}{2^73^2\Gamma(j_1)(j_1+3/2)}\frac{1}{(x-y)_+^2(x-y)_-^{2j_1+4}},
\end{eqnarray}
where in the second line we introduced new variables \(s_1=s\alpha \ \ s_2=s(1-\alpha)\).

Similarly we can get integral representation in case of three-point correlators (\ref{3G2d})-(\ref{3S2d}):

\begin{eqnarray}
B^g_{j_1j_2j_3}&=&2^{-7}b(j_1,j_2,j_3)
\int\limits_{-1}^1\int\limits_{-1}^1d\alpha d\beta(1-\alpha^2)^2(1-\beta^2)^2C^{\frac{5}{2}}_{j_1-1}(\alpha)C^{\frac{5}{2}}_{j_3-1}(\beta)L^{\frac{5}{2}}_{j_2-1},\label{BGInt}\\
B^q_{j_1j_2j_3}&=&-2b(j_1,j_2,j_3)
\int\limits_{-1}^1\int\limits_{-1}^1d\alpha d\beta(1-\alpha^2)(1-\beta^2)C^{\frac{3}{2}}_{j_1}(\alpha)C^{\frac{3}{2}}_{j_3}(\beta)L^{\frac{3}{2}}_{j_2},\label{BQInt}\\
B^s_{j_1j_2j_3}&=&2^{2}b(j_1,j_2,j_3)
\int\limits_{-1}^1\int\limits_{-1}^1d\alpha d\beta C^{\frac{1}{2}}_{j_1+1}(\alpha)C^{\frac{1}{2}}_{j_3+1}(\beta)L^{\frac{1}{2}}_{j_2+1},\label{BSInt}
\end{eqnarray}
where
\begin{gather}
b(j_1,j_2,j_3)=i^{j_1+j_2+j_3+3}2^{j_1-j_2+j_3}\mathcal{N}^3(N_c^2-1)\sigma_{j_1}\sigma_{j_2}\sigma_{j_3}(j_1+j_2+j_3+5)!,\\
L^{\kappa}_n=C^{\kappa}_{n}(\frac{1-\alpha \beta}{\alpha-\beta})\frac{(1+\alpha)^{j_3+2}(\alpha-\beta)^{n}(1+\beta)^{j_1+2}}{(2+\alpha+\beta)^{j_1+j_2+j_3+6}}.
\end{gather}

Due to the fact that \((1-\alpha^2)^{\kappa-\frac{1}{2}}\) is a measure on interval \((-1,1)\) for Gegenbauer polynomials \(C^{\kappa}_{n}(\alpha)\)(\ref{orthonorm_Gegenbauer}) we can interpret expressions (\ref{BGInt})-(\ref{BSInt}) as an projection of \(L^{\kappa}_n\) on Gegenbauer polynomials.

\par\medskip

\section{Three-point correlator of operators with spins.}\label{AppCorWithSpins}

This appendix is a reminder of the formulas obtained in  the paper \cite{Costa:2011dw}, with some precisions for our particular cases.  According to its methods, a formula for correlation function of any three primary operators with dimensions \(\Delta_i\) and spins \(l_i\) was obtained, using the embedding formalism. Below we give their expression in original notations and apply it to our particular case of twist-2 operators. Embedding formalism implies the embedding of physical space \(\mathcal{V}=\mathcal{R}^d\) (\(\mathcal{R}^{d-k,k}\)) into the space
\(\mathcal{M}=\mathcal{R}^{1,d+1}\) (\(\mathcal{R}^{d-k+1,k+1}\)) where the conformal group \(SO(1,d+1)\) (\(SO(d-k+1,k+1)\)) is realized linearly. The vector \(x\) from \(\mathcal{V}\) lifts up to \(\mathcal{M}\) by the formula \(x \leftrightarrow P_x=(1,x^2,x)\) , which sets the one-to-one correspondence of vectors from \(\mathcal{V}\) and light-rays in \(\mathcal{M}\). Scalar product of two vectors \(P_1=(P_{1+},P_{1-},p_1)\) and \(P_2\) from \(\mathcal{M}\) sets as \((P_1\cdot P_2)=-\frac{P_{1+}P_{2-}+P_{1-}P_{2+}}{2}+p_1 p_2\), where   \(p_1 p_2\) means the scalar product in \(\mathcal{V}\). In the paper \cite{Costa:2011dw},   three vectors of polarization \(Z_i \leftrightarrow z_i\) were introduced which contract  tensor indices of each operator: \(\phi(x,z)=\phi_{a_1,...,a_l}z^{a_1}...z^{a_l}\). In our case this corresponds to the projection of all indexes on \(n_+\) direction as in \ref{O^gg}-\ref{O^ss}. Thus in our case all indices have the same polarization \(z_1=z_2=z_3=n_+\). The formula for three-point correlation function reads in these notations as follows:

\begin{gather}
\langle\Phi(P_1,Z_{n_+})\Phi(P_2,Z_{n_+})\Phi(P_3,Z_{n_+})\rangle=\sum\limits_{n_{12},n_{13},n_{23}\geq 0}\lambda_{n_{12},n_{13},n_{23}}
\begin{bmatrix}
\Delta_1 & \Delta_2 & \Delta_3\\
l_1 & l_2 & l_3 \\
n_{23}& n_{13}&n_{12}
\end{bmatrix},\label{SumTens}
\end{gather}
where summation goes over all possible tensor structures. The coefficients \(\lambda_{n_{12},n_{13},n_{23}}\) are labeled by the set \(\{n_{12},n_{13},n_{23}\}\) of integers satisfying the following inequalities \(m_1=l_{1}-n_{12}-n_{13}\geq 0 \), \(m_2=l_{2}-n_{12}-n_{23}\geq 0 \), \(m_3=l_{3}-n_{13}-n_{23}\geq 0 \) and the tensor structures are explicitly given by

\begin{gather}
\begin{bmatrix}
\Delta_1 & \Delta_2 & \Delta_3\\
l_1 & l_2 & l_3 \\
n_{23}& n_{13}&n_{12}
\end{bmatrix}=
\frac{V_1^{m_1}V_2^{m_2}V_3^{m_3}H_{12}^{n_{12}}H_{13}^{n_{13}}H_{23}^{n_{23}}}{P_{12}^{\frac{1}{2}(\tau_1+\tau_2-\tau_3)}P_{13}^{\frac{1}{2}(\tau_1+\tau_3-\tau_2)}
P_{23}^{\frac{1}{2}(\tau_2+\tau_3-\tau_1)}},
\end{gather}
where
\begin{gather}
\tau_i=\Delta_i+l_i,\\
P_{ij}=-2(P_i\cdot P_j)=x_{ij}^2,\\
H_{ij}=-2\left((Z_i\cdot Z_j)(P_i\cdot P_{j})-(Z_i\cdot P_j)(Z_j\cdot P_{i})\right)=-2x_{ij+}^2,\\
V_{i,jk}=\frac{(Z_i\cdot P_j)(P_i\cdot P_k)-(Z_i\cdot P_k)(P_i\cdot P_j)}{(P_j\cdot P_k)},\\
V_1=V_{1,23}=\frac{x_{21+}x_{13}^2-x_{31+}x_{12}^2}{x_{23}^2}, \\
V_2=V_{2,31}=\frac{x_{32+}x_{21}^2-x_{12+}x_{23}^2}{x_{13}^2}, \\
V_3=V_{3,12}=\frac{x_{13+}x_{23}^2-x_{23+}x_{13}^2}{x_{12}^2}.
\end{gather}

It is easy to see that in the general case all tensor structures are different. In the case  when the coordinates are restricted to the \(\{n_+,n_-\}\) - plane we get for \(V\)  much simpler expressions

\begin{gather}
V_1=-\frac{x_{12+}x_{13+}}{x_{23+}},\ V_2=-\frac{x_{23+}x_{12+}}{x_{13+}},\ V_3=-\frac{x_{13+}x_{23+}}{x_{12+}}.
\end{gather}
In the case of twist-2 operators with spin \(j+1\) it reduces to

\begin{gather}
\begin{bmatrix}
j_1+1 & j_2+3 & j_3+3\\
j_1+1 & j_2+1 & j_3+1 \\
n_{23}& n_{13}&n_{12}
\end{bmatrix}=\frac{(-1)^{j_1+j_2+j_3-n_{12}-n_{13}-n_{23}+3} 2^{n_{12}+n_{13}+n_{23}-j_1-j_2-j_3-6}}{x_{12+}x_{13+}x_{23+}x_{12-}^{j_1+j_2-j_3+2}x_{13-}^{j_1+j_3-j_2+2}x_{23-}^{j_2+j_3-j_1+2}}.
\end{gather}
As one can see in this case, all tensor structures have the same coordinate dependence and thus we can rewrite \ref{SumTens}  as follows:

\begin{gather}
\langle O_{j_1}O_{j_2}O_{j_3}\rangle=\frac{C_{j_1j_2j_3}}{x_{12+}x_{13+}x_{23+}x_{12-}^{j_1+j_2-j_3+2}x_{13-}^{j_1+j_3-j_2+2}x_{23-}^{j_2+j_3-j_1+2}}.
\label{3pRC}
\end{gather}
For the case of two twist-2 operators \(O_{j_1}(x_1)\), \(O_{j_2}(x_2)\) with spins \(j_1+1\), \(j_2+1\) and one Konishi \(O_K(x_3)\) with spin \(0\) and bare dimension \(4\) we get

\begin{gather}
\langle O_{j_1}(x_1)O_{j_2}(x_2)O_K(x_3)\rangle=\frac{C_{j_1j_2}}{x_{13+}^2x_{23+}^2x_{12-}^{j_1+j_2+2}x_{13-}^{j_1-j_2+2}x_{23-}^{j_2-j_1+2}}. \label{T2T2Kspinstr}
\end{gather}

\section{Diagrams at $g^2$ order}\label{App1loop}
\subsection{One scalar-scalar-gluon vertex}

Here we present all terms which appear in
\begin{gather}
\langle\Tr(X(\alpha_1)X(\alpha_2)A_+(\alpha_3))\Tr(Z(\beta_1)Z(\beta_2))\Tr(\bar{X}(\gamma_1)\bar{X}(\gamma_2)\bar{Z}(\gamma_3)\bar{Z}(\gamma_4))\Phi_{ssg}\rangle,
\end{gather}
where \(\Phi_{ssg}=\int d^4u\Tr((\partial^\mu\bar{X}A_{\mu}X-\bar{X}A_{\mu}\partial^{\mu}X)+(\partial^\mu XA_{\mu}\bar{X}-XA_{\mu}\partial^{\mu}\bar{X}))(u)\) is one scalar-scalar-gluon vertex.
The terms corresponding to Fig.\ref{ris:1gl-s-s V}.a. read as follows:

\begin{gather}
\int d^4u \partial_{\beta_2+}\left((-2N_c^2+6-\frac{4}{N_c^2})F_{2a1}+
(N_c^4-5N_c^2+8-\frac{4}{N_c^2})F_{2a2}\right)+(\beta_1\leftrightarrow \beta_2),\\
-\int d^4u \partial_{\gamma_4+}\left((-2N_c^2+6-\frac{4}{N_c^2})F_{2a1}+
(N_c^4-5N_c^2+8-\frac{4}{N_c^2})F_{2a2}\right)+(\beta_1\leftrightarrow \beta_2),\\
\int d^4u \partial_{\gamma_4+}\left((4-\frac{4}{N_c^2})F_{2a1}+
(-2N_c^2+6-\frac{4}{N_c^2})F_{2a2}\right)+(\beta_1\leftrightarrow \beta_2),\\
-\int d^4u \partial_{\beta_2+}\left((4-\frac{4}{N_c^2})F_{2a1}+
(-2N_c^2+6-\frac{4}{N_c^2})F_{2a2}\right)+(\beta_1\leftrightarrow \beta_2).
\end{gather}

where
\begin{eqnarray}
F_{2a1}&=&F(\alpha_1,\alpha_2,\alpha_3,\beta_1,\beta_2,\gamma_1,\gamma_2,\gamma_3,\gamma_4)=\notag\\
&=&\frac{1}{|\alpha_3-u|^2|\gamma_3-\beta_1|^2|\gamma_4-u|^2|\beta_2-u|^2|\gamma_1-\alpha_1|^2|\gamma_2-\alpha_2|^2},\\
F_{2a2}&=&F(\alpha_2,\alpha_1,\alpha_3,\beta_1,\beta_2,\gamma_1,\gamma_2,\gamma_3,\gamma_4).
\end{eqnarray}

For Fig.\ref{ris:1gl-s-s V}.b. we get:

\begin{gather}
\int d^4u \partial_{\beta_2+}\left((4-\frac{4}{N_c^2})F_{2b1}+
(-2N_c^2+6-\frac{4}{N_c^2})F_{2b2}\right)+(\beta_1\leftrightarrow \beta_2),\\
-\int d^4u \partial_{\gamma_3+}\left((4-\frac{4}{N_c^2})F_{2b1}+
(-2N_c^2+6-\frac{4}{N_c^2})F_{2b2}\right)+(\beta_1\leftrightarrow \beta_2),\\
\int d^4u \partial_{\gamma_3+}\left((-2N_c^2+6-\frac{4}{N_c^2})F_{2b1}+
(N_c^4-5N_c^2+8-\frac{4}{N_c^2})F_{2b2}\right)+(\beta_1\leftrightarrow \beta_2),\\
\int d^4u \partial_{\beta_2+}\left((-2N_c^2+6-\frac{4}{N_c^2})F_{2b1}+
-(N_c^4-5N_c^2+8-\frac{4}{N_c^2})F_{2b2}\right)+(\beta_1\leftrightarrow \beta_2),
\end{gather}

where
\begin{eqnarray}
F_{2b1}&=&F(\alpha_1,\alpha_2,\alpha_3,\beta_1,\beta_2,\gamma_1,\gamma_2,\gamma_4,\gamma_3),\\
F_{2b2}&=&F(\alpha_2,\alpha_1,\alpha_3,\beta_1,\beta_2,\gamma_1,\gamma_2,\gamma_4,\gamma_3).
\end{eqnarray}
Summing up all this terms we get zero in the limit, when \(\gamma_i\rightarrow \gamma\).

\par\medskip

For the second Konishi term \(\Tr\bar{X}\bar{Z}\bar{X}\bar{Z}\) we get the same expressions with replacement \(\gamma_2\leftrightarrow \gamma_3\) and the same color factor \((-N_c^2+5-\frac{4}{N_c^2})\) for all terms, which again leads to the full cancellation.
\subsection{4-scalar vertex}\label{1loopDiag}

Direct calculation of the contribution from \(\Tr 2Z\bar{X}\bar{Z}X\) gives us

\begin{gather}
\langle\Tr\bar{X}(\gamma_1)\bar{X}(\gamma_2)\bar{Z}(\gamma_3)\bar{Z}(\gamma_4)\Tr{X(\alpha_1)X(\alpha_2)}\Tr Z(\beta_1)Z(\beta_2)\int d^4u\Tr (2Z\bar{X}\bar{Z}X)(u) \rangle=\notag\\
=2\mathcal{N}^6\left((-N_c^2+4-\frac{3}{N_c^2})\frac{T(\gamma_2,\gamma_3,\alpha_2,\beta_2)}{|\gamma_1-\alpha_1|^2|\beta_1-\gamma_4|^2}+(-N_c^2+4-\frac{3}{N_c^2})\frac{T(\gamma_1,\gamma_3,\alpha_2,\beta_2)}{|\gamma_2-\alpha_1|^2|\beta_1-\gamma_4|^2}+\right. \notag\\
\left.+(-N_c^2+4-\frac{3}{N_c^2})\frac{T(\gamma_2,\gamma_4,\alpha_2,\beta_2)}{|\gamma_1-\alpha_1|^2|\beta_1-\gamma_3|^2}+(-N_c^2+4-\frac{3}{N_c^2})\frac{T(\gamma_1,\gamma_4,\alpha_2,\beta_2)}{|\gamma_2-\alpha_1|^2|\beta_1-\gamma_3|^2}\right)
+\notag\\
+\left(\begin{smallmatrix}\alpha_1 \leftrightarrow \alpha_2,\\ \beta_1\leftrightarrow \beta_2
\end{smallmatrix}\right),\label{ZbXbZX}
\end{gather}
where we have introduced function
\begin{gather}
T(\gamma_i,\gamma_j,\alpha,\beta)=\int d^4\omega\frac{1}{|\gamma_i-\omega|^2|\gamma_j-\omega|^2|\alpha-\omega|^2|\beta-\omega|^2}. \label{DefT}
\end{gather}
Euclidian version of this function \(T_E=-iT\) has  the asymptotic form at \(\gamma_i,\gamma_j\to\gamma\)  as follows:
\begin{gather}
T_E(\gamma_i,\gamma_j,\alpha,\beta)\simeq\frac{\pi^2}{|\alpha-\gamma|^2|\beta-\gamma|^2}(2+\log\frac{|\alpha-\gamma|^2|\beta-\gamma|^2}{\epsilon_{ij}^2|\alpha-\beta|^2}), \ \ \epsilon_{ij}=\gamma_i-\gamma_j \label{Teucl}
\end{gather}
which can be obtained from the exact expression for (\ref{DefT}) which was obtained in \cite{Usyukina:1992jd}. Due to this asymptotic form all terms in \ref{ZbXbZX} have the same coordinate dependence \(\frac{\pi^2(2+\log\frac{|\alpha_2-\gamma|^2|\beta_2-\gamma|^2}{|\alpha_2-\beta_2|^2|\epsilon|^2})}{|\alpha_1-\gamma|^2|\alpha_2-\gamma|^2|\beta_1-\gamma|^2|\beta_2-\gamma|^2}\). We can omit the  \(2\) in the numerator because it gives us a term which cancels by action of \(\mathcal{G}^{\frac{1}{2};g=0}_{j_1+1,\alpha_1,\alpha_2}\mathcal{G}^{\frac{1}{2};g=0}_{j_2+1,\beta_1,\beta_2}\) as in \(g^0 \)   case. Moreover , due to the same reason we can retain only the overall scale of  \(\epsilon_{ij}\)'s in \ref{Teucl}, because the change of this scale only leads to an extra term \(\frac{const}{|\alpha_1-\gamma|^2|\alpha_2-\gamma|^2|\beta_1-\gamma|^2|\beta_2-\gamma|^2}\), disappearing after the action of derivatives.
Collecting all coefficients we get for the contribution from \(\Tr 2Z\bar{X}\bar{Z}X\) to the Konishi term \(-2\Tr(\bar{X}\bar{X}\bar{Z}\bar{Z})(\gamma)\)

\begin{gather}
4g^22\cdot4(-N_c^2+4-\frac{3}{N_c^2})\Omega,
\end{gather}
where the  function \(\Omega\) is given by
\begin{gather}\Omega=\frac{\mathcal{N}^6\pi^2}{|\alpha_1-\gamma|^2|\alpha_2-\gamma|^2|\beta_1-\gamma|^2|\beta_2-\gamma|^2}\log\frac{|\alpha_2-\gamma|^2|\beta_2-\gamma|^2}{|\alpha_2-\beta_2|^2|\epsilon|^2}+
\left(\begin{smallmatrix}\alpha_1 \leftrightarrow \alpha_2,\\ \beta_1\leftrightarrow \beta_2
\end{smallmatrix}\right).
\end{gather}
For the remaining five terms from \eqref{4-s.v.} we get :

\begin{gather}
4g^22\cdot4(-N_c^2+4-\frac{3}{N_c^2})\Omega,\\
-4g^2(2(-N_c^2+4-\frac{3}{N_c^2})+(N_c^4-4N_c^2+6-\frac{3}{N_c^2})+(N_c^2+2-\frac{3}{N_c^2}))\Omega,\\
-4g^2(2(-N_c^2+4-\frac{3}{N_c^2})+(N_c^4-4N_c^2+6-\frac{3}{N_c^2})+(N_c^2+2-\frac{3}{N_c^2}))\Omega,\\
-4g^2(2(-N_c^2+4-\frac{3}{N_c^2})+(N_c^4-4N_c^2+6-\frac{3}{N_c^2})+(N_c^2+2-\frac{3}{N_c^2}))\Omega,\\
-4g^2(2(-N_c^2+4-\frac{3}{N_c^2})+(N_c^4-4N_c^2+6-\frac{3}{N_c^2})+(N_c^2+2-\frac{3}{N_c^2}))\Omega.
\end{gather}

Carrying out a similar calculation for the Konishi term \(2\Tr\bar{X}\bar{Z}\bar{X}\bar{Z}\) we get:

\begin{gather}
-4g^22\cdot2(2(N_c^4-4N_c^2+6-\frac{3}{N_c^2})+2(N_c^2+2-\frac{3}{N_c^2}))\Omega,\\
-4g^22\cdot2(2(N_c^4-4N_c^2+6-\frac{3}{N_c^2})+2(N_c^2+2-\frac{3}{N_c^2}))\Omega,\\
4g^24(-N_c^2+4-\frac{3}{N_c^2})\Omega,\\
4g^24(-N_c^2+4-\frac{3}{N_c^2})\Omega,\\
4g^24(-N_c^2+4-\frac{3}{N_c^2})\Omega,\\
4g^24(-N_c^2+4-\frac{3}{N_c^2})\Omega.
\end{gather}

And finally, summing up all terms we obtain

\begin{eqnarray}
&&\langle\Tr X(\alpha_1)X(\alpha_2)\Tr Z(\beta_1)Z(\beta_2)\Tr[\bar{X},\bar{Z}]^2(\gamma)\rangle=\notag\\
&&=-48\pi^2g^2\mathcal{N}^6(N_c^4-N_c^2)\frac{1}{|\alpha_1-\gamma|^2|\alpha_2-\gamma|^2|\beta_1-\gamma|^2|\beta_2-\gamma|^2}\log\frac{|\alpha_2-\gamma|^2|\beta_2-\gamma|^2}{|\alpha_2-\beta_2|^2|\epsilon|^2}+\nn\\
&&\hskip10cm +\left(\begin{smallmatrix}\alpha_1 \leftrightarrow \alpha_2\\
\beta_1\leftrightarrow \beta_2
\end{smallmatrix}\right).\label{4sclContrib}
\end{eqnarray}

\printindex

\end{document}